\documentclass[twocolumn]{pasj01}
\Received{$\langle$reception date$\rangle$}
\Accepted{$\langle$acception date$\rangle$}
\Published{$\langle$publication date$\rangle$}
\SetRunningHead{Astronomical Society of Japan}{Usage of \texttt{pasj00.cls}}

\usepackage{graphicx}
\usepackage{bm}
\usepackage{ulem}
\usepackage[dvips]{color}
\usepackage{xspace}
\usepackage{lscape}

\SetRunningHead{Astronomical Society of Japan}{Usage of \texttt{pasj00.cls}}

\renewcommand{\figurename}{Fig.}
\renewcommand{\tablename}{Table}
\renewcommand\thefootnote{\alpha{footnote}}

\newcommand{\lx}{$L_{\rm x}$}

\begin{document}
\title{\textcolor{black}{Hard} X-ray Luminosity Function of Tidal Disruption Events:
First Results from {\it MAXI} Extragalactic Survey}

\author{
 Taiki Kawamuro\altaffilmark{1},
 Yoshihiro Ueda\altaffilmark{1},
 Megumi Shidatsu\altaffilmark{2},
 Takafumi Hori\altaffilmark{1},
 Nobuyuki Kawai\altaffilmark{3},
 Hitoshi Negoro \altaffilmark{4}, and
 Tatehiro Mihara \altaffilmark{2}. 
}

\Received{} 
\Accepted{}

\altaffiltext{1}{Department of Astronomy, Kyoto University, Kyoto 606-8502, Japan}
\altaffiltext{2}{{\it MAXI} team, RIKEN, 2-1, Hirosawa, Wako-shi, Saitama 351-0198, Japan}
\altaffiltext{3}{Department of Physics, Faculty of Science
Tokyo Institute of Technology, Tokyo 152-8551, Japan}
\altaffiltext{4}{Department of Physics, Nihon University, 1-8-14 Kanda-Surugadai, Chiyoda-ku, Tokyo 101-8308, Japan}
\email{kawamuro@kusastro.kyoto-u.ac.jp}

\KeyWords{galaxies: individuals ({\it Swift} J1112.2-8238, {\it Swift} J164449.3+573451, 
{\it Swift} J2058.4+0516, NGC 4845) -- X-rays: galaxies}
\maketitle

\begin{abstract} 
We derive the first \textcolor{black}{hard} X-ray luminosity function (XLF) of 
stellar tidal disruption events (TDEs) by supermassive black holes (SMBHs), 
which gives an occurrence rate of TDEs per unit volume as a function of peak 
luminosity and redshift, utilizing an unbiased sample observed by the
Monitor of All-sky X-ray Image ({\it MAXI}). 
On the basis of the light curves characterized by a power-law decay with an 
index of $-5/3$, a systematic search using the {\it MAXI} data in the first 37 months 
detected four TDEs, all of which have been found in the literature. 
To formulate the TDE XLF, we consider the mass function of SMBHs, that of 
disrupted stars, the specific TDE rate as a function of SMBH mass, and the 
fraction of TDEs with relativistic jets. We perform an unbinned maximum 
likelihood fit to the {\it MAXI} TDE list and check the consistency
with the observed TDE rate in the {\it ROSAT} all sky survey.  
The results suggest that the intrinsic fraction of the jet-accompanying 
events is $0.0007$--$34\%$. We confirm that at 
$z \lesssim 1.5$ the contamination by TDEs to the hard X-ray
luminosity functions of active galactic nuclei is not significant and
hence that their contribution to the growth of SMBHs is negligible at the redshifts.
\end{abstract}

\section{INTRODUCTION}\label{intro}

The nature of supermassive black holes (SMBHs) that reside in 
inactive galaxy nuclei is very difficult to explore, compared with 
those of accreting ones observed as active galactic nuclei (AGNs). 
However, when the orbital path of a star is close enough to a SMBH to be
disrupted by the tidal force exceeding the self-gravity of the star,
a luminous flare in the UV/X-ray bands is predicted \citep{Rees88}.
This is called a tidal disruption event (TDE). Observations of TDEs are 
important to take a census of dormant SMBHs and to investigate their environments.
Moreover, thanks to their large luminosities, TDEs provide us with valuable 
opportunities to study distant ``inactive'' galactic nuclei.

X-ray surveys covering a large sky area are very useful to detect TDEs, 
because we cannot predict when and where an event occurs. 
\textcolor{black}{
In fact, wide-area X-ray surveys performed with {\it ROSAT}, {\it XMM-Newton}, 
{\it INTEGRAL}, and {\it Swift} (e.g., \cite{Kom99}; \cite{Esq07};
\cite{Bur11}; \cite{Sax12}; \cite{Niko13}) have discovered many of the TDEs 
reported so far.  Some of TDEs have also been detected in the optical and UV 
bands (e.g., \cite{Gez06}, 2008, 2012; \cite{Arc14}; \cite{Hol14}; \cite{Van11}). 
Since the first detection of a TDE, which occurred in NGC 5905 \citep{Bade96}, 
a few tens of X-ray TDEs have been identified \citep{Kom12}. 
}
The identifications of TDEs were mainly based on their variability characteristics,
such as a large amplitude and the unique decline law of the light curve 
(e.g., \cite{Kom99}), which are supported by both analytic solutions (\cite{Rees88}; 
\cite{Phi89}) and numerical simulations (e.g., \cite{Eva89}).

The recent hard X-ray survey with {\it Swift}/BAT and subsequent X-ray
observations detected three TDEs accompanied by relativistic
jets (\cite{Bur11}; \cite{Cen12}; \cite{Bro15}). Presence of the jets 
was suggested also from follow-up observations in the radio band 
\citep{Zau11}. In these TDEs, the X-ray fluxes were dominated by 
non-thermal emission in the beamed jets, unlike in ``classical'' TDEs, 
where one observes blackbody radiation emitted from the stellar debris 
accreted onto the SMBH. Thus, it is interesting to explore what fraction 
of TDEs produces relativistic jets and what the physical mechanism 
to launch the jets is.

Theoretically, the occurrence rate of TDEs is estimated to be $10^{-5}$--$10^{-4}$ 
galaxy$^{-1}$ yr$^{-1}$ (e.g., \cite{Mag99}), and its dependence on SMBH 
mass is calculated (e.g., \cite{Wang04}; \cite{Sto14}). In fact, many
observational results (e.g., \cite{Don02}; \cite{Esq08}; \cite{Mak10}) 
are in rough agreements with the predicted TDE rate. An important
quantity that describes the statistical properties of TDEs is the
``luminosity function'', i.e., the luminosity dependence of the TDE
rate. The luminosity function of TDEs is highly useful in evaluating the
effect of TDEs on the growth history of SMBHs and in predicting the number 
of detectable events in future surveys. Considering that the flare
luminosity of a TDE depends on the SMBH mass (e.g., \cite{Ulm99};
\cite{Li02}), it is possible theoretically to estimate the luminosity
function of TDEs \citep{Milo06}. However, observational studies that
directly constrain the TDE luminosity function based on a statistically
complete sample have been highly limited so far.

In this paper, we derive the \textcolor{black}{hard} X-ray luminosity function 
of TDEs, using a statistically complete sample obtained with the Monitor of 
All-sky X-ray Image ({\it MAXI}) mission. For this purpose, we systematically 
search for hard X-ray transient events at high galactic latitudes $(|b| >10^\circ)$, 
and identify TDEs. We then derive the luminosity functions of TDEs associated 
with and without relativistic jets individually. 
This result also enables us to estimate the contribution of TDEs to the growth 
of SMBHs.

This paper is organized as follows. Section \ref{maxi_overview} presents
the overview of the {\it MAXI} observations. In Section \ref{lc_ana},
the light-curve analysis of the {\it MAXI} sources to identify TDEs is
presented. The derivation of the X-ray luminosity function of TDEs is
described in Section~\ref{ml_ana}. Section~\ref{dis} gives discussion 
based on our XLF model, including the contribution of TDEs to the XLF 
of AGNs and to the evolution of the SMBH mass density. Section~\ref{sum} 
presents the summary of our work. Appendix~\ref{app:sec:ta_search} 
describes the method of detecting transient sources from the {\it MAXI} 
data. Throughout this paper, we assume a $\Lambda$ cold dark-matter
model with $H_0 = 70$ km s$^{-1}$ Mpc$^{-1}$, $\Omega_{\rm M}$ = 0.3,
and $\Omega_\Lambda$ = 0.7.  The ``$\log$'' denotes the base-10
logarithm, while the ``$\ln$'', the natural logarithm.

\section{Search of TDE From {\it MAXI} Data}\label{ta_search}

\subsection{Observations and Data Reduction}\label{maxi_overview}

The {\it MAXI} mission \citep{Mat09} on the International Space Station
(ISS) has been monitoring all sky in the X-ray band since 2009. {\it MAXI}
achieves the highest sensitivity as an all-sky monitor, and is highly 
useful to detect X-ray transient events, including TDEs. It carries two
types of cameras, the Gas Slit Cameras (GSCs; \cite{Mih11}), consisting of 12 counters,
and the Solid-state Slit Camera \citep{Tom11}. In this paper, we only utilize
the data of the GSCs, which covers the energy band of 2--30 keV. The GSCs
have two instantaneous fields-of-view of 1$^\circ$.5$\times$160$^\circ$
separated by 90 degrees. They rotate with a period of 92 minutes
according to the orbital motion of the ISS, and eventually covers a
large fraction of the sky (95\%) in one day \citep{Sug11}.


To search the {\it MAXI}/GSC data for transient events, we analyze those
taken in the first 37 months since the beginning of the operation (from 2009
September 23 to 2012 October 15). We also restrict our analysis to high
galactic latitudes ($|b|>10^\circ$). Exactly the same data were analyzed to 
produce the second {\it MAXI}/GSC catalog  \citep{Hiroi13}, which 
contains 500 sources detected in the 4--10 keV band from the data
integrated over the whole period. The details of the data selection
criteria are described in Section~2 of \citet{Hiroi13}.

\subsection{Identification of TDEs in {\it MAXI} Catalogs}\label{id_tde}

TDEs are transient events that become bright for a typical time scale of
months to years (e.g., \cite{Kom99}). Hence, they may be missed with the {\it MAXI} Alert
System \citep{Neg12}, which is currently optimized to detect variability of
sources on time scales from hours to a few days. Also, faint TDEs may
not be detected in the 2nd {\it MAXI} catalog \citep{Hiroi13}, because
the long integration time of 37 months works to smear the signals,
making the time averaged significance lower than the threshold.

To detect such transient events as completely as possible, we newly
construct the {\it MAXI}/GSC ``{\it transient source catalog}'' based on the same
data as used by \citet{Hiroi13}. The analysis is optimized to find variable objects
on the time scale of 30 or 90 days. Namely, we split the whole data into
30 or 90 day bins, and independently perform source detection from each
dataset to search for new sources that are not listed in the 2nd {\it
MAXI} catalog \citep{Hiroi13}. As a result, we detect 
10 transient sources with the detection significance $s_{\rm D} > 5.5$ in either of
the time-sliced datasets, where $s_{\rm D}$ is defined as (best-fit flux
in 4-10 keV) / (its 1$\sigma$ statistical error). The details of the
analysis procedure and the resultant light curves of the transient
sources are given in Appendix~\ref{app:sec:ta_search}. As an example, Figure \ref{detection_ex}
shows the {\it MAXI}/GSC significance map around {\it Swift} J164449.3+573451
(hereafter {\it Swift} J1644+57), a transient source detected by this
method. This object is significantly detected in the data of 30 days
during the outburst (left panel), while it is not in the 37-month data
(right panel). We estimate the sensitivity limit for the peak flux
averaged for 30 days is $\sim 2.5$ mCrab.

\begin{figure*}
\hspace{1cm}
\includegraphics[scale=0.32]{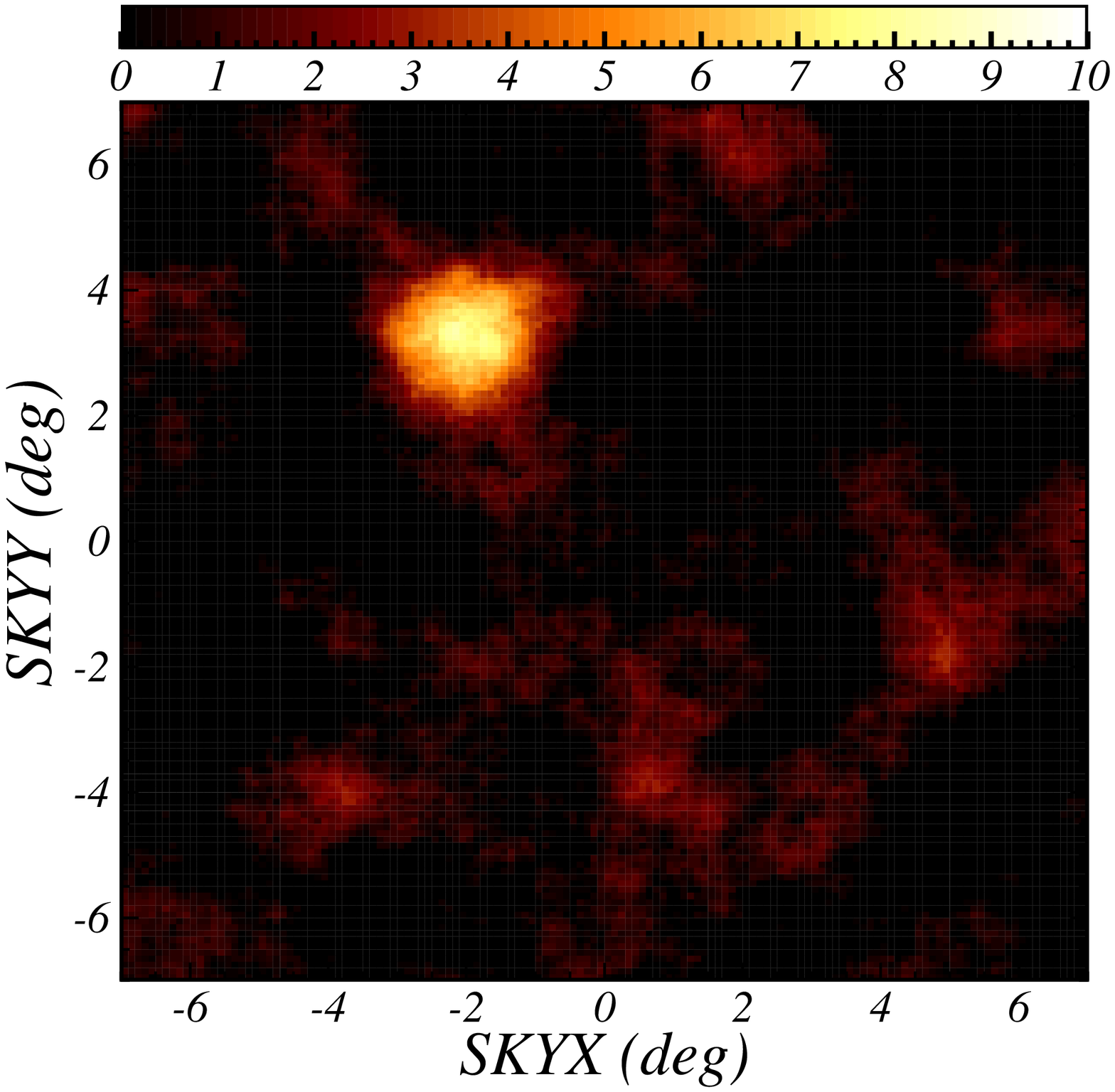} \hspace{2cm}
\includegraphics[scale=0.32]{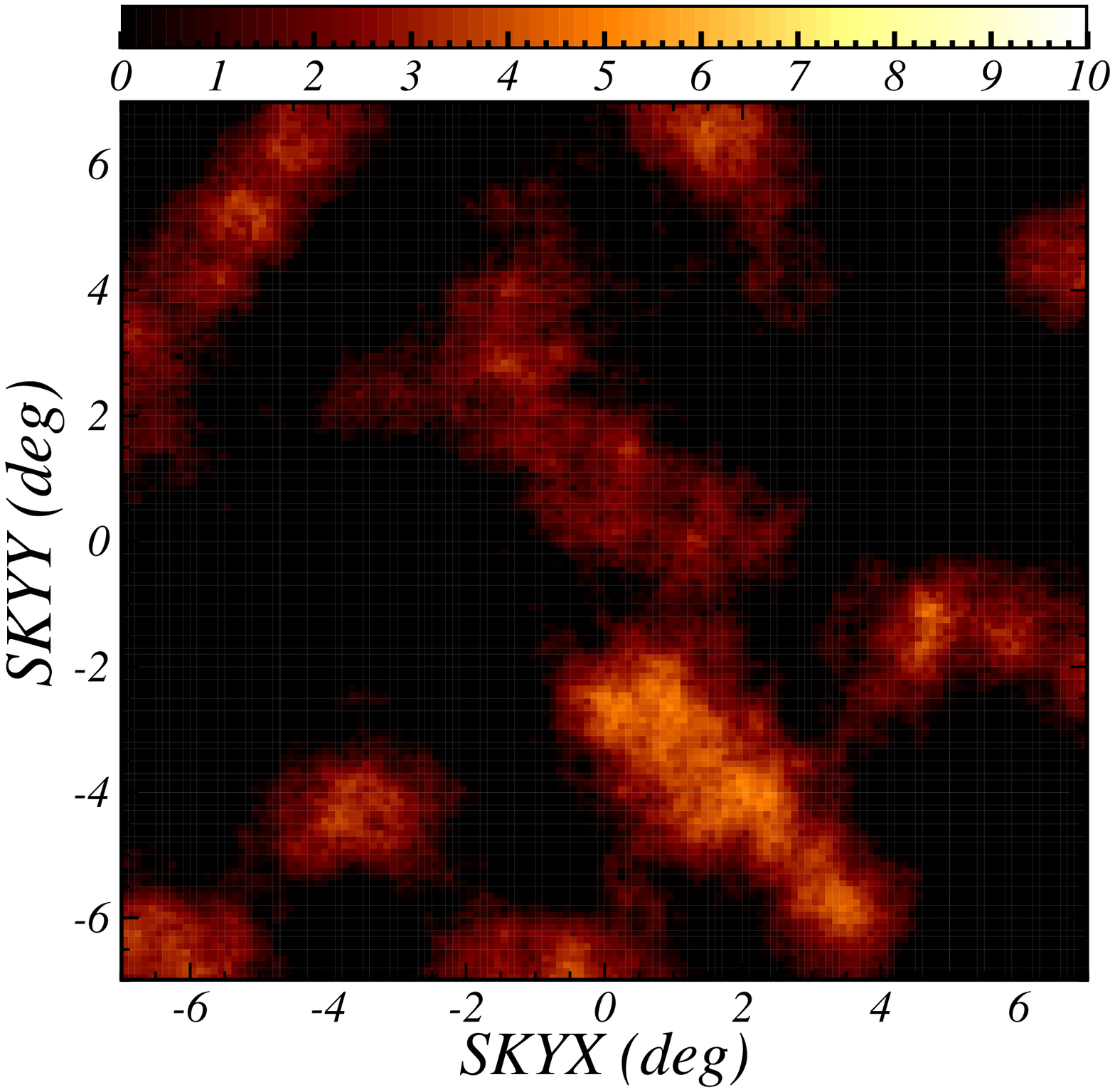}
\caption{
(Left) Significant map around {\it Swift} J1644+58 (left upper source) obtained from
the data integrated for 30 days when the object was the brightest.
(Right) The same but obtained from the total 37-month data.
}
\label{detection_ex}
\end{figure*}


On the basis of positional coincidence, we identify 3 TDE \textcolor{black}{candidates} 
in the {\it MAXI} transient source catalog from the literature, {\it Swift} 
J1112.2-8238 (hereafter {\it Swift} J1112-82; \cite{Bro15}), {\it Swift} 
J1644+57 \citep{Bur11}, and {\it Swift} J2058.4+0516 (hereafter {\it Swift} 
J2058+05; \cite{Cen12}). They are located at $z= 0.89$, $z= 0.354$, and $z= 1.1853$, 
and the (isotropic) luminosities in the 4--10 keV band are estimated to be 
10$^{47.1}$ erg s$^{-1}$, 10$^{46.6}$ erg s$^{-1}$, and 10$^{47.5}$ erg s$^{-1}$, 
respectively.  We also identify another \textcolor{black}{candidate} that occurred in 
NGC 4845 \citep{Niko13}, which has been already listed in the 2nd {\it MAXI}/GSC 
catalog \citep{Hiroi13}. The 30-day averaged peak flux in the 4--10 keV band is 
3.2 mCrab, with a significance of $s_{\rm D} = 6.8$, which corresponds to a luminosity 
of 10$^{42.3}$ erg s$^{-1}$ at $z= 0.004110$. 
\textcolor{black}{
We note, however, that the X-ray flare of NGC 4845 may be attributed to
a variable AGN, which is supported by the radio observation of the
unresolved central core \citep{Irw15}. From the time evolution of
the radio spectrum, \citet{Irw15} also suggested the presence of an
expanding outflow or a jet, which may be associated with the X-ray
flare. It is not yet unclear whether the X-ray flare is due to a TDE or
the AGN.}  The detailed information of each TDE from the literature
(\cite{Bur11}; \cite{Cen12}; \cite{Niko13}; \cite{Zau13}; \cite{Bro15};
\cite{Pas15}) is summarized in Table \ref{tde_sample}.

\begin{table*}
\caption{Our Sample of Tidal Disruption Events\label{tde_sample}}
\begin{center}
\begin{tabular}{lccccc}
\hline
Name & $z$ & $\log L_{4-10 {\rm keV}}$ & $\Gamma$ & $\delta$ & $M_\ast$ \\ 
$[1]$ & [2] & [3]  & [4] & [5] & [6]   \\ \hline 
 {\it Swift} J1112.2-8238     & 0.89     & 47.1  & -    & -  &  -   \\
 {\it Swift} J164449.3+573451 & 0.354    & 46.6  & 10   & 16 & 0.15 \\
 {\it Swift} J2058.4+0516     & 1.1853   & 47.5  & $>2$ & -  & 0.1 \\
 NGC 4845                     & 0.004110 & 42.3  & -    & -  & 0.02 \\ 
\hline
\multicolumn{1}{@{}l@{}}{\hbox to 0pt{\parbox{160mm}
{\footnotesize
\textbf{Notes.}\\
Col. [1]: Name of the TDE.  \\ 
Col. [2]: Redshift.  \\
Col. [3]: Luminosity (erg s$^{-1}$) in the 4--10 keV band.  \\
Col. [4]: The Bulk Lorentz factor.  \\
Col. [5]: The Doppler factor. \\
Col. [6]: \textcolor{black}{Accreted mass in units of solar mass.} \\
}
\hss}}
\end{tabular}
\end{center}
\end{table*}

\subsection{Search for Unidentified TDEs}\label{lc_ana}

\begin{figure*}[!ht]
\includegraphics[scale=0.6,angle=-90]{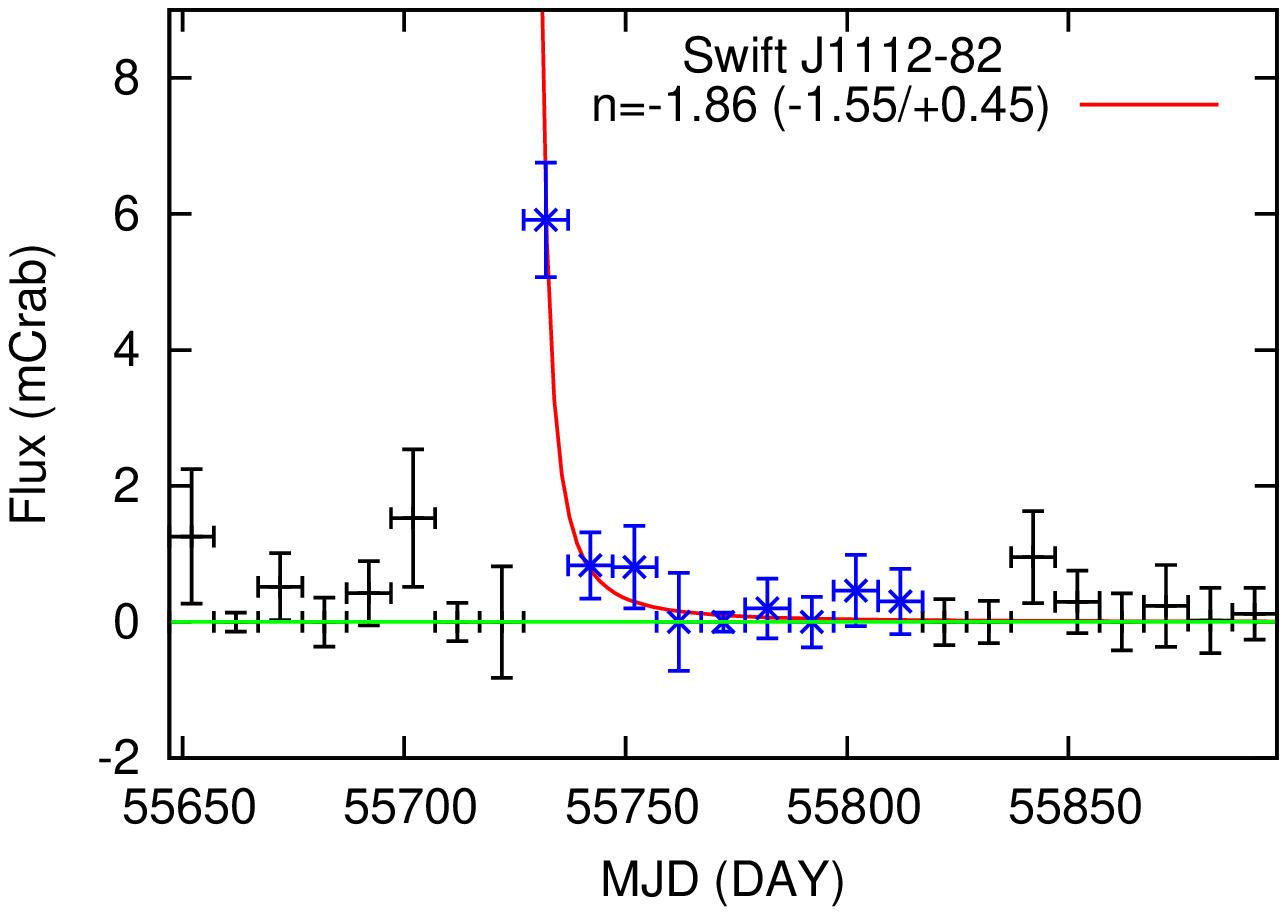} \hspace{0.5cm}
\includegraphics[scale=0.6,angle=-90]{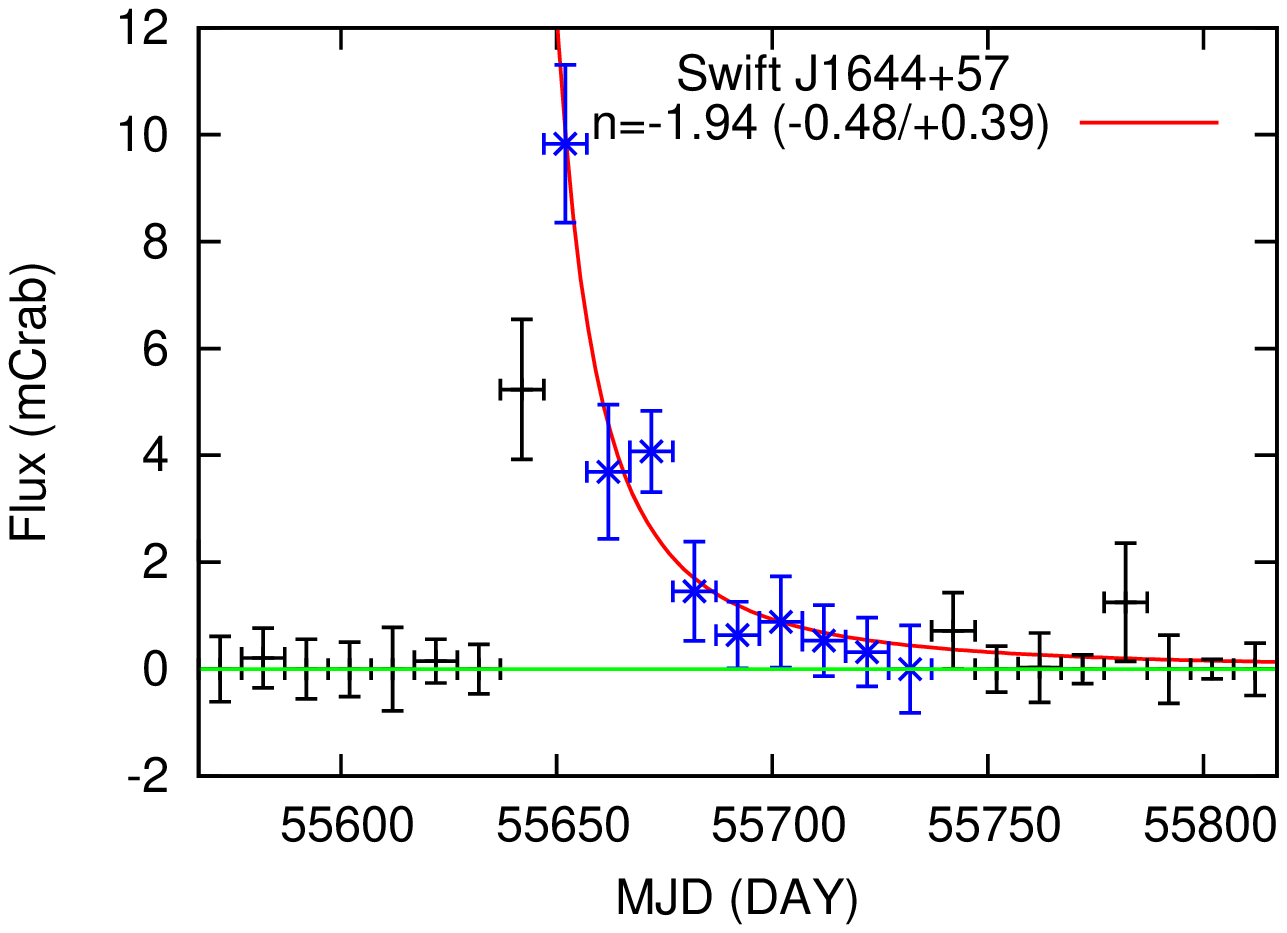}\\
\includegraphics[scale=0.6,angle=-90]{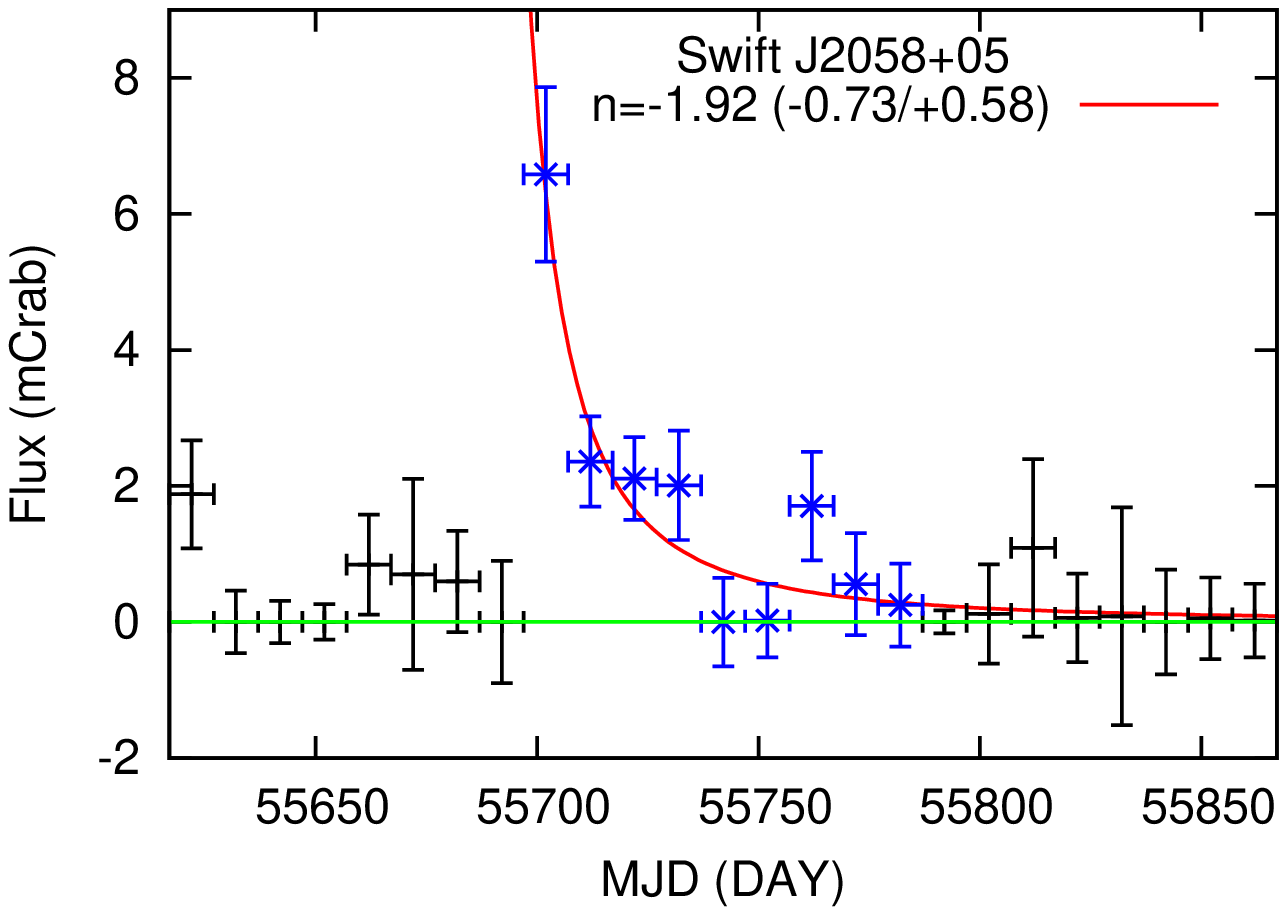} \hspace{0.5cm}
\includegraphics[scale=0.6,angle=-90]{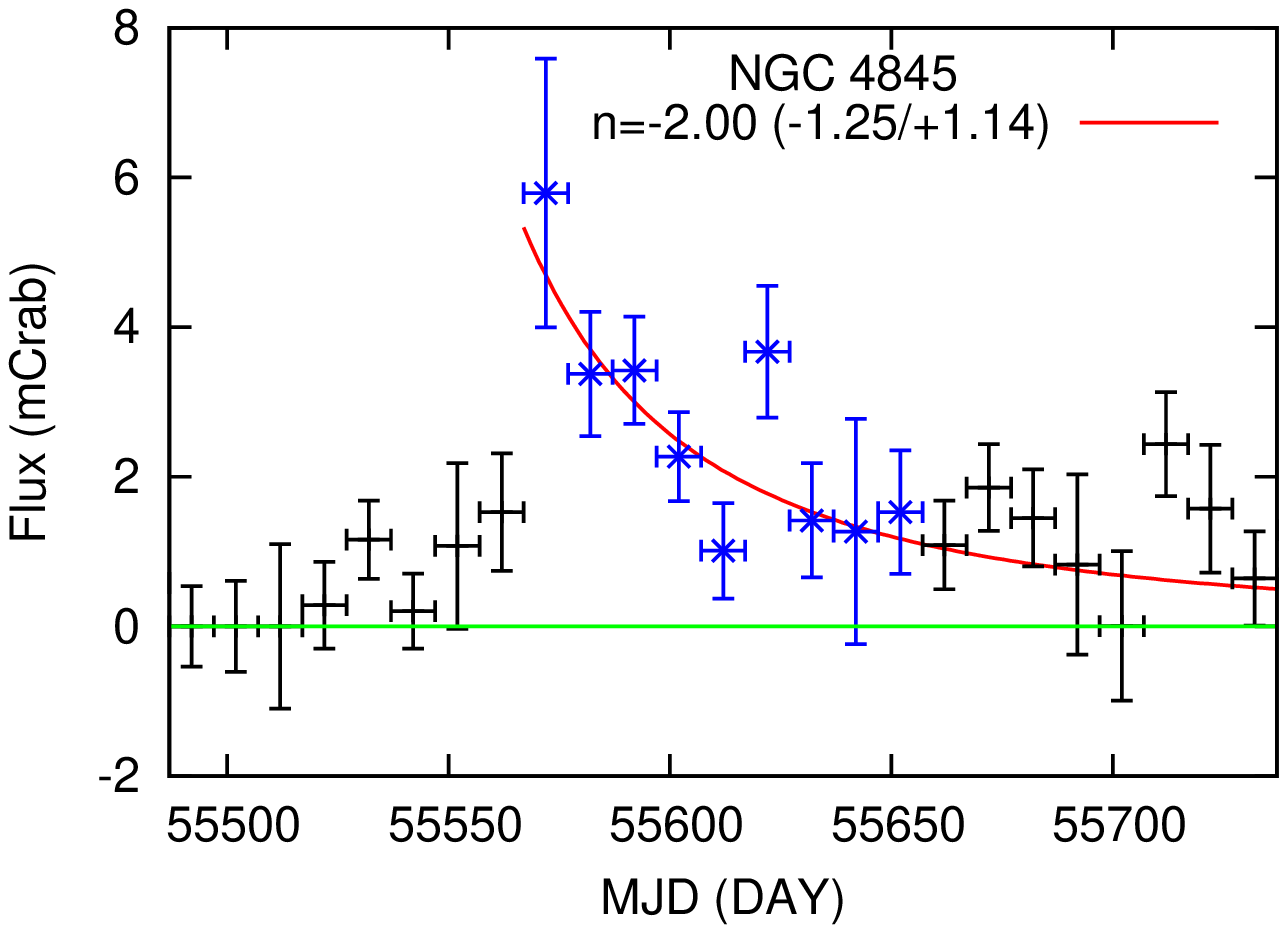} 
\caption{
X-ray (3--10 keV) light curves of four TDEs during their flares.
Only blue regions are used to be fit with a power-law decay model. 
The best-fit power-law index (n) is indicated in each panel with 
errors at 90\% confidence level.
}
\label{tde_lc}
\end{figure*}



\begin{figure*}[!ht]
\includegraphics[scale=0.40,angle=-90]{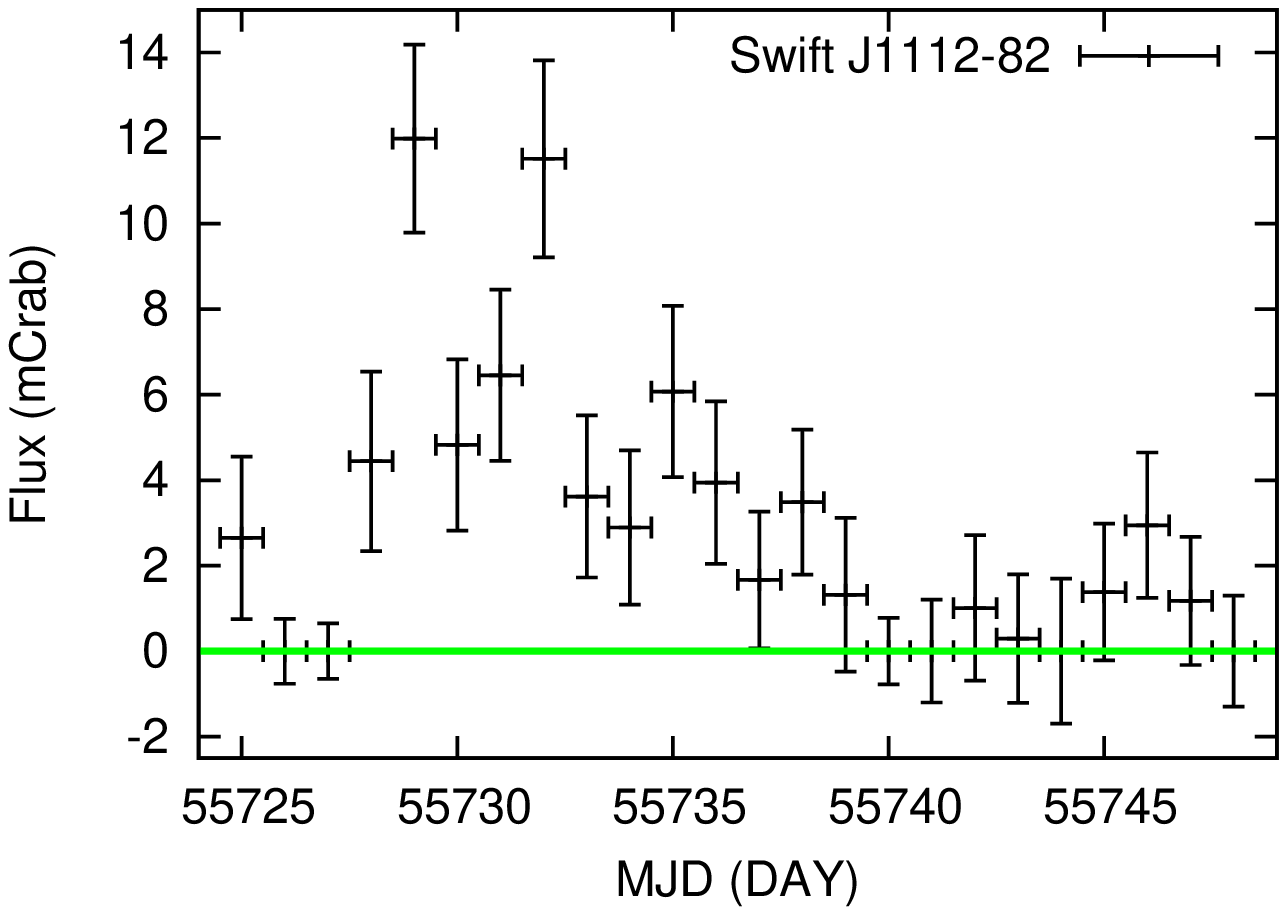} 
\includegraphics[scale=0.40,angle=-90]{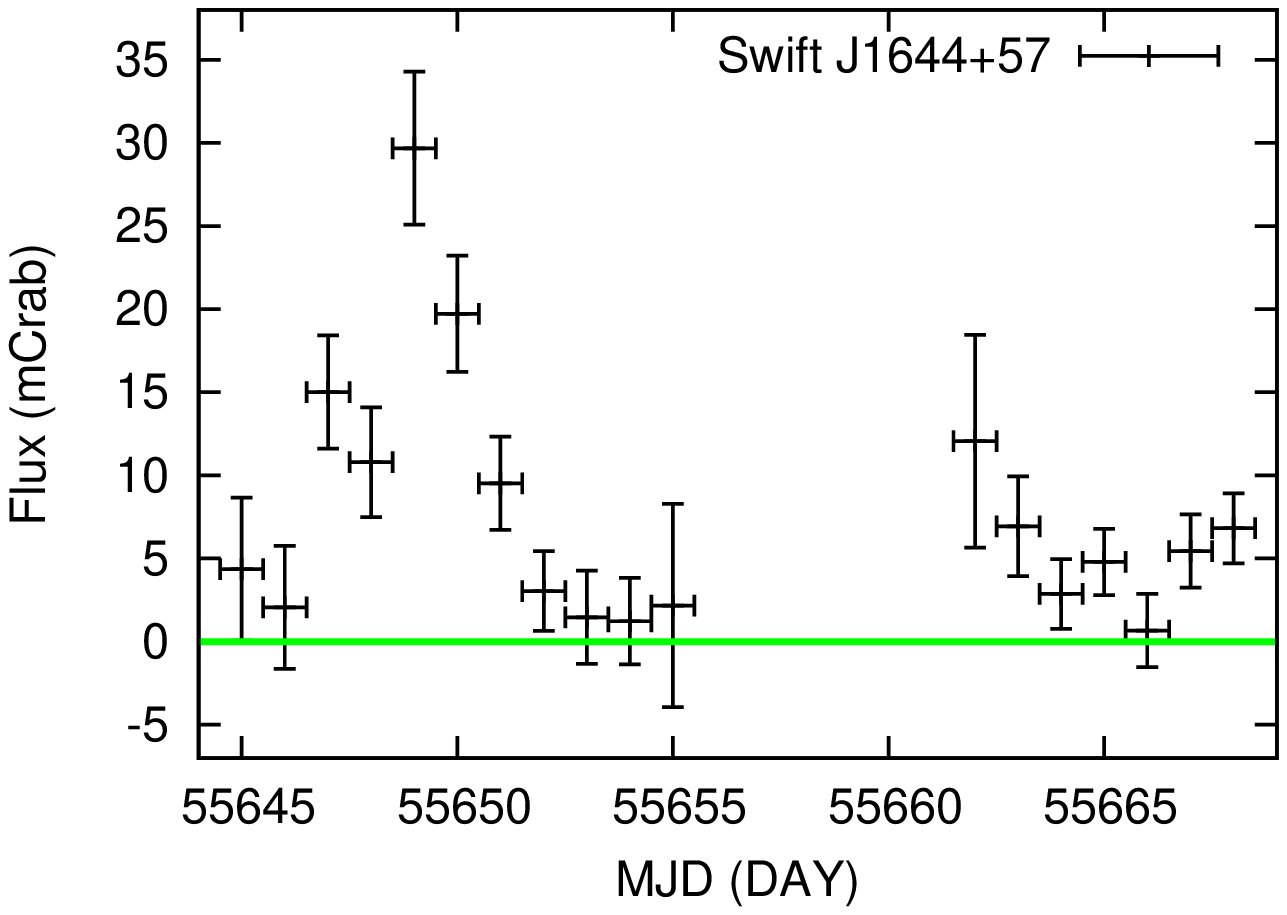}
\includegraphics[scale=0.40,angle=-90]{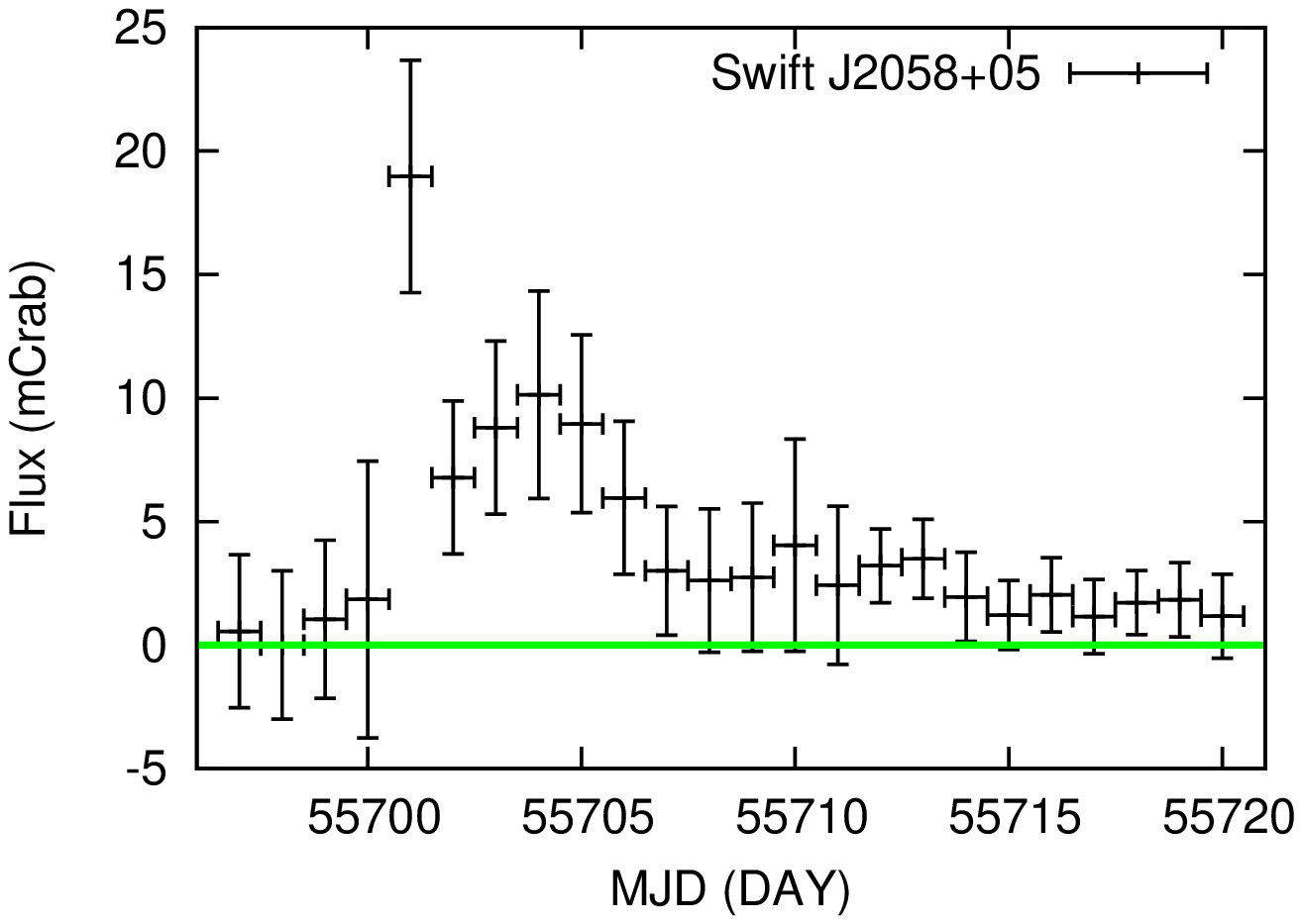}
\caption{
X-ray (3--10 keV) light curves of 
the three TDEs with relativistic jets around their peak luminosities in 1 day bins.
For {\it Swift} J1644+57 
no {\it MAXI} data were obtained around MJD $\sim$55660.
}
\label{tde_lc_2}
\end{figure*}

To constrain statistical properties of TDEs, such as the occurrence rate
as a function of X-ray luminosity (i.e., X-ray luminosity function = XLF
hereafter), it is very important to perform their complete survey at a
given flux limit. Hence, we search for other possible TDEs that are not
reported in the literature from these {\it MAXI} catalogs. We make use of the
general characteristics of the light curve pattern of TDEs. As mentioned
in Section~\ref{intro}, a TDE shows a rapid flux increase followed by a power-law
decay with an index of -5/3 as a function of time (\cite{Rees88}; \cite{Phi89}).

For this purpose, we make the light curves of all sources in the 2nd
{\it MAXI}/GSC catalog \citep{Hiroi13} and in the transient catalog 
(Appendix~\ref{app:sec:ta_search}) in 10 days, 30 days, and 90 days bins, 
in three energy bands, 3--4 keV, 4--10 keV, and 3--10 keV. The fluxes in 
each time bin are obtained by the same image fitting method as described 
in Appendix~\ref{app:sec:ta_search} by fixing the source
positions. We discard the data when the photon statistics is too poor
within each bin (see subsection 2.1 in \citet{Iso15} for details). Using 
the light curve of Crab nebula analyzed in the same way, we estimate a
systematic uncertainty in the flux is $\sim$10\%, which is added to the
statistical error.

As the reference, we analyze the light-curve pattern of the four identified
TDEs, {\it Swift} J1112-82, {\it Swift} J1644+57, {\it Swift} J2058+05, and 
NGC 4845. We find that all of them show the following two characteristics. 
The first one is high variability amplitudes; the ratio between the highest 
flux and the one of the previous bin in the 30-day (90-day) averaged light 
curves is 7.0 (1.6), 16.3 (7.7), 8.2 (2.4), and 1.4 (6.8) for  {\it Swift} 
J1112-82, {\it Swift} J1644+57, {\it Swift} J2058+05, and NGC 4845, respectively. 
Here, we assign a flux of 5$\times10^{-12}$ erg s$^{-1}$ cm$^{-2}$ (4--10 keV) 
for each bin when the source is undetected, which is the sensitivity limit of 
the 2nd {\it MAXI} catalog \citep{Hiroi13}. The second characteristic is that 
the decay light curves are consistent with a power-law profile of $t^{-5/3}$,
where $t$ is time since the onset time of each TDE. We note that the time of 
the flux peak, $t_{\rm p}$, is delayed from the TDE onset time by approximately 
80 days for NGC 4845, 20 days for {\it Swift} J1644+57 and {\it Swift} J2058+05, 
and 5 days for {\it Swift} J1112-82.
Hence, we set the central day of the bin showing the highest flux 
as $t_{\rm p}$, and estimate the TDE onset time by correcting for these offsets.  
We confirm that the 10-days light curves in the 3--10 keV band follow power-law 
profiles, as shown in Figure~\ref{tde_lc}. A power-law fit to the light curve over 
90 days after the peak flux is found to be acceptable in terms of a $\chi^2$ test,
yielding the best-fit index of $-1.86^{+0.45}_{-1.55}$, $-1.94^{+0.39}_{-0.48}$, 
$-1.92^{+0.58}_{-0.73}$, and $-2.00^{+1.14}_{-1.25}$ for {\it Swift} J1112-82, 
{\it Swift} J1644+57, {\it Swift} J2058+05, and NGC 4845, respectively. The errors
denote statistical ones at 90\% confidence limits, and they are all consistent 
with $-5/3$. Accordingly, we apply the above two conditions to the light curves of
all {\it MAXI} sources (in total 506) except for the four TDEs. First, we find  12 
(12) objects satisfy  
the criterion that the ratio between the highest and second highest flux
bins is larger than 5 in the 30 (90) day light curve. For these
candidates, we then perform the same light-curve fitting with a
power-law profile as described above. The time delay from the TDE onset
to the observed flux peak is set to be either 5 days, 20 days, or 80
days. As a result, we find that none of them show a decaying index
consistent with $-5/3$ except for the objects identified as AGNs or
X-ray galactic sources.  Thus, we conclude that {\it MAXI} detected only
the four TDEs identified above during the first 37 months of its
operation, which can be regarded as a statistically complete sample at
the sensitivity limit of {\it MAXI} for transient events, {\it as long as
TDEs share similar characteristics in the X-ray light curve to those of
the known events.} 
\textcolor{black}{ 
According to numerical simulations, the index of the power-law decay
becomes steeper than $-5/3$ when the star is not fully disrupted
\citep{Gui13}. Such events would be missed in our sample.  
We find that four TDE candidates reported by \citet{Hry16} from the {\it
Swift}/BAT ultra-hard X-ray band (20--195 keV) data were not
significantly detected in the {\it MAXI} data covering the same epoch. The
details are given in Appendix~\ref{app:sec:bat_src}. In all events, the
flux upper limit in the {\it MAXI} band is smaller by a factor of $\sim$4
than that expected from the Swift/BAT flux by assuming a photon index of
2.0. This implies that these TDE candidates might have unexpectedly hard
spectra or be subject to heavy obscuration. Such TDEs, if any, are not
considered in our analysis. }

The continuous monitoring data of {\it MAXI} provide us with unique 
information on the X-ray light curve of TDEs in the 3--10 keV band, which 
can be compared with those obtained with {\it Swift}/BAT in the 14--195 
keV band. Figure~\ref{tde_lc_2} plots the {\it MAXI} light curves in 1-day 
bins of the three TDEs except for NGC 4845, which was too faint to be examined
on shorter time scales than 10 days. We find that the luminosity peaked at 
MJD = 55729, MJD = 55649, and MJD = 55701 for {\it Swift} J1112--82, {\it Swift} 
J2058+05, and {\it Swift} J1644+57, respectively, which are all consistent 
with those determined with {\it Swift}/BAT within 1 day (\cite{Bur11}; 
\cite{Cen12}; \cite{Bro15}). Thus, there is no evidence for significant 
($>1$ day) time lags between the soft ($< 10$ keV) and hard ($> 10$ keV) 
bands.

\section{\textcolor{black}{Hard} X-ray Luminosity Function of Tidal Disruption Events}
\label{ml_ana}

\subsection{Definition of Luminosities}\label{l_def}

\begin{figure}[htb]
\includegraphics[scale=0.23]{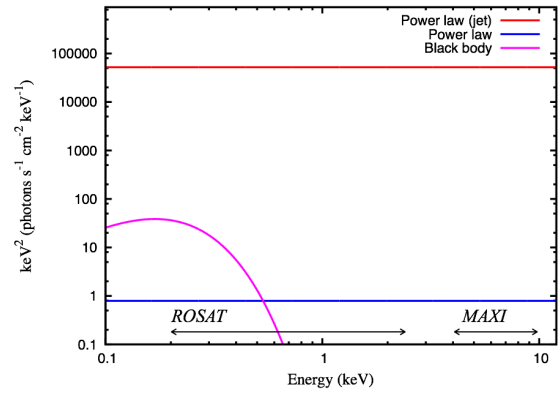}
\caption{
A representative spectrum of a TDE with $T_{\rm bb} = 5\times10^{5}$ K. 
The three components are shown (red: jet component with $\delta = 16$,
blue: Comptonized component, magenta: blackbody component). 
The unit of the vertical axis is arbitrary.
}
\label{fig:tde_sed}
\end{figure}

In this section, we summarize the definitions of TDE luminosities used 
in our analysis ($L^{\rm obs}_{\rm X,ins} (t)$, $L^{\rm obs}_{\rm X}$, and \lx ). 
To make luminosity conversion between different energy bands, we need to 
assume model spectra of TDEs. According to the previous studies on TDEs 
observed in the X-ray band (e.g., \cite{Esq07}; \cite{Bur11}; \cite{Niko13}), 
we can approximate that the X-ray spectra of TDEs without jets are composed of
blackbody radiation and a power law, which originate from the optically thick disk and 
its Comptonized component by hot corona, respectively. For TDEs with jets, 
a relativistically beamed power law is added to the above spectrum. A 
representative spectrum is shown in Figure~\ref{fig:tde_sed}, where the photon 
indices of the two power-law components are assumed to be 2. Modelling the 
decay profile of the lightcurve of a TDE with $(t/t_{\rm p})^{-5/3}$, we can 
relate an observed, instantaneous luminosity at $t$ in a given energy band 
calculated by assuming isotropic emission, $L^{\rm obs}_{\rm X,ins} (t)$, to 
its corresponding peak luminosity $L^{\rm obs}_{\rm X}$, as 
\begin{small}
\begin{eqnarray}
 L^{\rm obs}_{\rm X,ins} (t) =  L^{\rm obs}_{\rm X} 
\Big(\frac{t}{t_{\rm p}}\Big)^{-5/3} (t \geq t_{\rm p}).
\end{eqnarray}
\end{small} 
We define \lx\ as the ``intrinsic'' peak luminosity of the Comptonized power-law
component in the rest-frame 4--10 keV band. It can be converted into the
``observed'' peak luminosity by
\begin{small}
\begin{eqnarray}
L^{\rm obs}_{\rm X} =  C L_{\rm X},
\end{eqnarray}
\end{small} 
where $C$ ($C_0$ or $C_1$; see below) is the conversion factor that depends on 
the shape of the spectrum, redshift, and viewing angle with respect to the jet 
axis (for TDEs with jets).

Introducing the fraction of TDEs with jets in total TDEs, $f_{\rm jet}$, 
we divide TDEs into two types, one with jets and the other without jets. For the 
TDEs without jets, whose fraction is $(1-f_{\rm jet})$, the conversion factor 
$C_0$ can be written as 
\begin{small}
\begin{eqnarray}
C_0 = \omega_{\rm pow} + \omega_{\rm bb},
\label{model_0}
\end{eqnarray}
\end{small}
where $\omega_{\rm pow}$ and $\omega_{\rm bb}$ are those 
for the Comptonized (power-law) and blackbody components, respectively.
We normalize $\omega_{\rm pow}$ to unity when 
$L^{\rm obs}_{\rm X}$ of a TDE at $z=0$ is 
defined in the 4--10 keV band and its spectrum is not absorbed.
The factor $\omega_{\rm bb}$ depends on the broadband spectrum
of a TDE, as detailed in the third paragraph of Section
\ref{sec:comp_rosat}. 
Note that the blackbody component is negligible in the 4--10 keV band
because its temperature is expected to be much lower than a few keV (Section~\ref{sec:comp_rosat}).

For TDEs with relativistic jets, the conversion factor $C_1$ can be written as
\begin{small}
\begin{eqnarray}
C_1 = \omega_{\rm pow} +  \omega_{\rm bb} + \omega_{\rm jet} \eta_{\rm jet} \delta^4
\label{model_1}
\end{eqnarray}
\end{small} 
where the last term represents the jet contribution. Here, 
$\omega_{\rm jet}$ takes account of the energy-band conversion, 
$\eta_{\rm jet}$ is the fraction of the intrinsic luminosity (i.e., 
that would be observed without beaming) of the jet in \lx\ 
(the peak luminosity of the Comptonized component), and 
$\delta$ is the Doppler factor. It is represented as
\begin{eqnarray}
\delta = \frac{1}{\Gamma - \sqrt{\Gamma^2-1} \cos \theta},
\end{eqnarray}
where $\Gamma$ is the Lorentz factor and $\theta$ is the viewing angle
with respect to the jet axis. The observed luminosity from the jet
becomes larger than the intrinsic one by a factor of $\delta^4$ with 
a frequency shift by $\delta$. In our analysis, we adopt $\Gamma=10$, 
which is suggested from the analysis of the spectrum energy distribution 
of {\it Swift} J1644+57 by \citet{Bur11}. Then, $\theta$ can be estimated 
if $\delta$ is constrained from the observations, as listed in 
Table~\ref{tde_sample} for {\it Swift} J1644+57. We also adopt $\eta_{\rm jet}=0.1$ 
as a standard value. It is confirmed that our main results do not sensitively 
depend on the choice of  $\Gamma$ or $\eta_{\rm jet}$ (see Section~\ref{sec:eff_par}).
\textcolor{black}{
Throughout our analysis, we do not take into account possible precession
of the jets due to the Lense-Thirring effect, as predicted by
\citet{Sto12}. This effect would be negligible because our calculation
of the XLF is based on the flux averaged over 30 days, which smears out the
effect. The conversion factors $C_0$ and $C_1$ are
dimensionless, composed only of the dimensionless factors ($\omega$,
$f_{\rm jet}$, and $\delta$).
}

\subsection{TDE Sample} \label{sec:tde_sample}

We regard that the four identified TDEs listed in Table~\ref{tde_sample}
constitute a complete sample from the {\it MAXI} survey for 37 months, and 
we utilize them to derive the XLF. When integrated for 30 days, the {\it MAXI} 
survey covers all the high Galactic latitudes ($|b|>10^\circ$) region, which 
corresponds to 83\% of the entire sky. As described in the previous section, 
we have searched for TDEs based on 30 days or 90 days binned light curves. Thus, 
the sensitivity limit for the 30-days averaged peak flux of TDEs to which our 
survey is complete is determined by that for 30-days integrated data of 
{\it MAXI}/GSC, which is 2.5 mCrab, or $3\times 10^{-11}$ erg s$^{-1}$ cm$^{-2}$ 
(4--10 keV). For simplicity, we ignore the dependence of the sensitivity on sky
positions \citep{Hiroi13}, whose effects are much smaller than the statistical 
uncertainties in the XLF parameters.

It is necessary to assume a spectrum to derive the luminosity from the
observed count rate of {\it MAXI}/GSC. Because the blackbody component can be 
ignored in the 4--10 keV band, here we assume a power-law spectrum absorbed with
the hydrogen column-densities reported by the previous studies (\cite{Bur11}; 
\cite{Cen12}; \cite{Niko13}; \cite{Bro15}). In the following analysis, for simplicity, 
we always adopt a photon index of 2 both for the Comptonization and jet components, 
as a representative value. In fact, for all four TDEs, the hardness ratio defined as 
$(H-S)/(H+S)$, where $H$ and $S$ is the count rates obtained from the {\it MAXI} data 
in the 4--10 keV and 3--4 keV bands, respectively, is consistent with a photon 
index of 2.0 within uncertainties. In this case, the K-correction factor is always 
unity and $\omega_{\rm pow} = \omega_{\rm jet}$. The calculated peak luminosity of each 
TDE is listed in Table~\ref{tde_sample}. Its systematic error due to the uncertainty 
in the peak-flux time within the bin size of the {\it MAXI} light curve 
has negligible effect on our conclusions. 


\subsection{Formulation of TDE X-ray Luminosity Function}

We define the XLF of TDEs so that \textcolor{black}{${\rm d} \Phi (L_{\rm x}, z)/{\rm d} L_{\rm x}$}
represents the TDE occurrence rate per unit co-moving volume per $L_{\rm x}$  per 
unit rest-frame time, as a function of $L_{\rm x}$ and $z$, in units of 
Mpc$^{-3}$ $L_{\rm x}^{-1}$ yr$^{-1}$. Note that this function has an additional dimension 
of per unit time compared with the ``instantaneous'' XLF of TDEs and the XLF of AGNs (see 
subsection \ref{cont_XLF}), which represents the number density of TDEs and AGNs,
respectively, {\it observed in a single epoch}.

In our work, we make simple assumptions for modelling the shape of the
TDE XLF. First, we write the TDE occurrence rate per unit volume as a
function of SMBH mass. It should be proportional to 
the product of the SMBH mass function (i.e., comoving number density of
SMBHs) and a specific TDE rate in a single SMBH. The local SMBH mass
function can be derived from the local luminosity function of galaxies by 
using the Faber-Jackson relation between the galaxy luminosity and the
SMBH mass, $L_{\rm gal} \propto M_{\rm BH}^k$ (e.g., \cite{Fer02};
\cite{Milo06}). It has a form of the Schechter function represented as 
\begin{eqnarray}
& \psi ( & M_{\rm BH \ast}; M_{\rm BH}) {\rm d}M_{\rm BH} \nonumber \\
& &  = \psi_0 \Big(\frac{M_{\rm BH}}{M_{{\rm BH}\ast}}\Big)^\gamma e^{-(\frac{M_{\rm BH}}{M_{{\rm BH}\ast}})^k}
\frac{{\rm d}M_{\rm BH}}{M_{{\rm BH} \ast}}.
\end{eqnarray}
Here, $\gamma = k (\alpha+1)-1$, where $\alpha$ is a parameter of the galaxy luminosity function 
defined as $\Psi (L_{\rm gal \ast}; L_{\rm gal}) {\rm d}L = 
\Psi_0 (L_{\rm gal}/L_{\rm gal \ast})^\alpha e^{-L_{\rm gal}/L_{\rm gal \ast}} 
{\rm d}L_{\rm gal}/L_{\rm gal \ast}$. The subscript $\ast$ indicates the characteristic parameter. 
Unless otherwise noted, we adopt $k = 0.8$, $\Psi_0 = 0.007$,  $\log (M_{\rm BH\ast}/M_\odot) = 
8.4$, and $\alpha = -1.3$ according to the results obtained by \citet{Mar03} and \citet{Bla01}, 
where $M_\odot$ is the solar mass. We refer to the dependence of the specific TDE rate on SMBH 
mass derived by \citet{Sto14}, 
\begin{eqnarray}
\xi \propto M_{\rm BH}^{\lambda}, 
\end{eqnarray}
where $\lambda$ is chosen to be $-0.4$.

To represent the TDE occurrence rate as a function of
{\it luminosity}, we further make an assumption that the peak
luminosity $L$ of a TDE (i.e., that free from the jet luminosity)
is proportional to the SMBH mass, or equivalently, a constant fraction 
of the Eddington luminosity, $\lambda_{\rm Edd}$. Then, by converting $M_{\rm BH}$ 
into $L$ in the product of $\psi(M_{\rm BH \ast}; M_{\rm BH})$ and $\xi$, we can  
express the occurrence rate of TDEs per unit volume in terms of $L$ as 
\begin{eqnarray}
\phi(L_\ast; L) {\rm d}L= \psi_0 \xi_0 \Big(\frac{L}{L_\ast}\Big)^{\gamma+\lambda}  
e^{-(\frac{L}{L_\ast})^k} \frac{{\rm d}L}{L_\ast}. 
\end{eqnarray}
We incorporate a redshift dependence of the TDE XLF with 
an evolution factor of $(1+z)^p$ that is multiplied to the local XLF.
Thus, the TDE XLF is formulated as 
\begin{eqnarray}
\frac{{\rm d} \Phi (L_{\rm x}, z)}{{\rm d} L_{\rm x}} {\rm d} L_{\rm x}  =  (1+z)^p \phi(L_{\rm x \ast}; L_{\rm x})  {\rm d}L_{\rm x}. 
\end{eqnarray}

\subsection{Mass Function of Stars Disrupted by SMBHs}

\textcolor{black}{ 
A TDE occurs only when the tidal disruption radius, 
$R_{\rm TDE} = R_\ast (M_{\rm BH}/M_\ast)^{1/3}$, where $M_{\ast}$ and $R_{\ast}$ are the 
mass and radius of the star, is larger than the Schwarzschild radius, $R_{\rm Sch}$ ($\equiv 2GM_{\rm BH}/c^2$, where $G$ is the gravitational constant.). 
For a given star, there is an upper boundary for the mass of a SMBH
that can cause a TDE.
Hence, the mass function of stars should be incorporated in calculating 
the actual TDE XLF.
}
We approximate it by the shape of an initial mass function (IMF) with an upper star mass 
boundary $M_{\ast, {\rm max}}$, considering that very massive stars are already dead 
due to their short life time. Here we employ the IMF derived by \citet{Kro01}, which
utilizes a larger sample than that by \citet{Sal55} and is similar to that by 
\citet{Chab03}. The normalized stellar mass function is continuously composed of next 
three equations; 
\begin{eqnarray}
&  P(M_{\ast}) & {\rm d}\log  M_\ast \propto  \nonumber \\ 
& &  M_\ast^{0.7}  (M_{\ast, {\rm min}} \leq M_\ast < 0.08 M_\odot) \\
& &   M_\ast^{-0.3}  (0.08 M_\odot \leq M_\ast < 0.5 M_\odot) \\
& &   M_\ast^{-1.3}  (0.5 M_\odot \leq M_\ast \leq M_{\ast, {\rm max}}), 
\end{eqnarray}
and satisfies   
\begin{equation}
\int^{M_{\ast, {\rm max}}}_{M_{\ast, {\rm min}}} P(M_\ast) {\rm d}\log M_\ast = 1,
\end{equation}
where we set $M_{\ast, {\rm min}} = 0.01 M_\odot$ and $M_{\ast, {\rm max}} = 1.0 M_\odot$. 
We confirm that the choice of $M_{\ast, {\rm max}}$ does not significantly affect our 
conclusions (Section~\ref{sec:eff_par}).

\subsection{Maximum Likelihood Fit}\label{sec:ml_func}

We adopt the unbinned maximum likelihood (ML) method to constrain the XLF parameters.
While the ML fit gives the best-fit parameters,
the goodness of the fit cannot be evaluated. Hence,
we perform one dimensional Kolmogorov-Smirnov test (hereafter KS test;
e.g., \cite{Pre92}) separately for the redshift distribution and for
the luminosity distribution between the observed data and best-fit
model. The p-value, the chance of getting observed data set, is
evaluated from the D-value assuming the one-sided KS test statistic. The
D-value is chosen to be the maximum value among the absolute distances
between an empirical cumulative distribution function and a theoretical
one.

We define the likelihood function as
\begin{scriptsize}
\begin{equation}
\mathcal{L} = -2 \sum_{i} \ln \frac{\int \int \int N(L_{\rm x}, L^{\rm obs}_{{\rm x} i}, z_i, M_\ast, \theta) {\rm d} L_{\rm x}  {\rm d}\log M_{\ast}{\rm d}\Omega/2\pi }
{\int \int\int\int \int N(L_{\rm x}, L^{\rm obs}_{\rm x}, z, M_\ast, \theta)  {\rm d} L_{\rm x} {\rm d} L^{\rm obs}_{\rm x} {\rm d}z {\rm d}\log M_{\ast}{\rm d}\Omega/2\pi },
\label{ml}
\end{equation}
\end{scriptsize}
where the subscript index $i$ refers to each TDE and the term 
$N(L_{\rm x}, L^{\rm obs}_{\rm x}, z, M_\ast, \theta)  {\rm d} L_{\rm x} {\rm d} 
L^{\rm obs}_{\rm x} {\rm d}z {\rm d}\log M_\ast {\rm d}\Omega / 2\pi $
represents the differential number of observable TDEs with the intrinsic peak luminosity 
$L_{\rm x}$, the observed one $L^{\rm obs}_{\rm x}$, the redshift $z$, the mass of the star 
$M_\ast$, and the viewing angle $\theta$, expected from the survey (note that $\theta$ is 
related to the solid angle as ${\rm d}\Omega = 2\pi {\rm d (cos}(\theta))$). 
By considering that the fraction of TDEs with jets among all TDEs is $f_{\rm jet}$, 
the differential number is calculated as

\begin{footnotesize}
\begin{eqnarray}
& N( & L_{\rm x}, L^{\rm obs}_{\rm x}, z, \theta, M_\ast)   \nonumber \\
&  & = \Bigl\{(1- f_{\rm jet})\delta_{\rm D} ( C_0 L_{\rm x } - L^{\rm obs}_{\rm x}  )  
       + f_{\rm jet} \delta_{\rm D} \Bigl(C_1 L_{\rm x } - L^{\rm obs}_{\rm x}  \Bigr) \Bigr\}  \nonumber \\
&  & \times \frac{{\rm d} \Phi (L_{\rm x}, z)}{{\rm d} L_{\rm x}} \frac{d_{\rm L}^2 (z)}{1+z} c\frac{{\rm d}\tau}{{\rm d}z} A
 \Big(\frac{L_{\rm x}^{\rm obs}}{d^2_{\rm L}}\Big) \frac{\Delta T}{1+z} P(M_\ast),
\end{eqnarray}
\end{footnotesize}
where $\delta_{\rm D}(x)$ is the Dirac $\delta$-function, $d_{\rm L}$ the luminosity 
distance, $c$ the light velocity, ${\rm d}\tau/{\rm d}z$ the differential look-back time, 
$A$ the survey area at the flux limit of $L_{\rm x}^{\rm obs}/4\pi
d^2_{\rm L}$, and $\Delta T$ ($=$ 37 months) the survey time at the observer's frame. 
The factor $1/(1+z)$ comes from the time dilation at $z$.

Since our sample size of TDEs is very small (four), we fix the following parameters 
of the XLF model, which cannot be well constrained from the data. We adopt the 
characteristic luminosity of $\log L_{\rm x \ast} = 44.6$ corresponding to the 
Eddington ratio of $\lambda_{\rm Edd} = 1$ (see next paragraph). As mentioned 
previously and listed in Table \ref{tab:def_par},  the Lorentz factor of the jets, 
the dependence of the specific TDE rate on SMBH mass, the fraction of the intrinsic 
jet luminosity in \lx, and the upper mass boundary of tidally disrupted stars are 
fixed at $\Gamma = 10$, $\lambda = -0.4$, $\eta_{\rm jet} = 0.1$, and 
$M_{\ast, {\rm max}} = 1.0 M_\odot$ as the standard parameters, respectively. 
Effects on the main results by changing these numbers ($\lambda_{\rm Edd}$, $\Gamma$, 
$\lambda$, $\eta_{\rm jet}$, and $M_{\ast, {\rm max}}$) from the default values will 
be examined in Section~\ref{sec:comp_rosat} and \ref{sec:eff_par}. The index of 
the redshift evolution is assumed to be either $p=0$ (no evolution case) or $4$ 
(strong evolution case). 
\textcolor{black}{
According to the prediction of numerical simulations 
that the occurrence rate of TDEs increases 
with the star-formation rate \citep{Aha15}, 
the latter case simply assumes that the TDE rate 
is proportional to the star-formation rate density, 
which evolves with $\propto (1+z)^4$ \citep{Per05}.
}
Eventually, only $\phi_0 \xi_0$ and $f_{\rm jet}$ are left as free parameters. 

The minimization procedure is carried out by using the MINUIT software package. 
We calculate the likelihood function over a redshift
range of $z =$ 0--1.5. The luminosity range is derived from the SMBH
range of $\log (M_{\rm BH}/M_\odot) =$ 4--8 for a given $\lambda_{\rm
Edd}$ (1.0). To convert the mass into the X-ray luminosity $L_{\rm x}$, we
consider a spectra composing of the three components described in Section~\ref{l_def} 
(the blackbody, Comptonization, and jet components). Specifically, we first determine 
the ratio of the Comptonized component in the 2--10 keV band to the bolometric 
luminosity without the jet component. Assuming that accretion physics in TDEs is 
similar to that of AGNs, we refer to the results by \citet{Vas07}, who derived the 
bolometric correction factor ($k_{\rm 2-10}$) from the 2--10 keV band to be $\sim$50 for 
$\lambda_{\rm Edd} = 1.0$. We then convert the luminosity in the 2--10 keV band to that 
in the 4--10 keV band by assuming a power-law photon index of 2.0. As a result, 
$\log L_{\rm x}$ spans a range from 40.2 to 44.2 for $\lambda_{\rm Edd} = 1.0$. The rest of the 
bolometric luminosity is attributed to the blackbody emission. The integration range 
of $M_\ast$ is determined to satisfy the criterion $R_{\rm TDE} / R_{\rm Sch} \geq 1 $. 
To calculate the tidal disruption radius  $R_{\rm TDE} = R_\ast (M_{\rm BH}/M_\ast)^{1/3}$, 
we convert the radius of a disrupted star $R_\ast$ to a mass $M_\ast$ with $R_\ast/R_\odot = 
(M_\ast/M_\odot)^{0.8}$ (e.g., \cite{Kipp90}). Hence, $R_{\rm TDE}/R_{\rm Sch}$ is a 
function of $M_\ast$ and $M_{\rm BH}$, or that of $M_\ast$, $\lambda_{\rm Edd}$, and $L_{\rm x}$.

Table~\ref{best_fit} summarizes the results of the ML fit for the 
cosmological evolution index of $p=0$ and $p=4$. 
The other fixed parameters are also listed in Table~\ref{tab:def_par}.
One-dimensional KS tests for the distribution of $z$ and for that of 
$L^{\rm obs}_{\rm x}$ do not rule out both results at the $90\%$ confidence 
level. The 90\% confidence upper and lower limits on $f_{\rm jet}$ are 
derived, corresponding to the case where the $\mathcal{L}$-value is 
increased by 2.7 from its minimum. Since the ML method cannot directly
determine the normalization ($\psi_0 \xi_0$) of the luminosity function,
we calculate it so that the predicted number from the model equals to
the detected number of the TDEs. The attached error corresponds to the 
Poisson error in the observed number at the 90\% confidence level 
based on equations (9) and (12) in \citet{Ge86}.

\begin{table}
\caption{Default Setting of Fixed Parameters\label{tab:def_par}}
\begin{center}
\begin{tabular}{cp{0.7cm}p{1.4cm}p{1.4cm}p{1.4cm}c}
\hline
$\lambda_{\rm Edd}$ &  $\Gamma$ & $\lambda$ &  $\eta_{\rm jet}$  &  $M_{\ast, {\rm max}}$ \\ 
  $[1]$ &  [2] & [3]  & [4] & [5]  \\ \hline 
1.0 &    10 & -0.4 &  0.1 &  1.0  \\ 
\hline
\multicolumn{1}{@{}l@{}}{\hbox to 0pt{\parbox{160mm}
{\footnotesize
\textbf{Notes.}\\
Col. [1]: Eddington ratio. \\
Col. [2]: The Lorentz factor. \\ 
Col. [3]: Index for the $M_{\rm BH}$ dependence of the TDE rate. \\
Col. [4]: Fraction of the intrinsic luminosity of the jet in $L_{\rm x}$. \\ 
Col. [5]: Upper mass boundary of disrupted stars in units of solar mass.  
}\hss}}
\end{tabular}
\end{center}
\end{table}

\begin{table}
\caption{Best-fit parameters\label{best_fit}}
\begin{center}
\begin{tabular}{ccccc}
\hline
p & $f_{\rm jet}$ &   $\psi_0\xi_{0}$ & p-value ($L^{\rm obs}_{\rm x}$-dist/$z$-dist)  \\ 
  $[1]$ & [2] & [3] & [4] \\ \hline 
0 & 0.012$^{+0.122}_{-0.010}$  & $1.7^{+1.6}_{-0.9}$  & 0.16/0.14  \\
4 & 0.003$^{+0.027}_{-0.002}$  & $1.6^{+1.6}_{-0.9}$  & 0.60/0.40 \\ 
\hline
\multicolumn{1}{@{}l@{}}{\hbox to 0pt{\parbox{160mm}
{\footnotesize
\textbf{Notes.}\\
Col. [1]: Index of the redshift evolution. \\
Col. [2]: Fraction of the TDEs with jets. \\
Col. [3]: Normalization factor of XLF in units of $10^{-8}$ Mpc$^{-3}$ $\log L_{\rm x}^{-1}$ yr$^{-1}$. \\
Col. [4]: p-value on the basis of the KS-test for each parameter of $L^{\rm obs}_{\rm x}$ and $z$. 
}\hss}}
\end{tabular}
\end{center}
\end{table}

\begin{figure}[htb]
\includegraphics[scale=0.35,angle=-90]{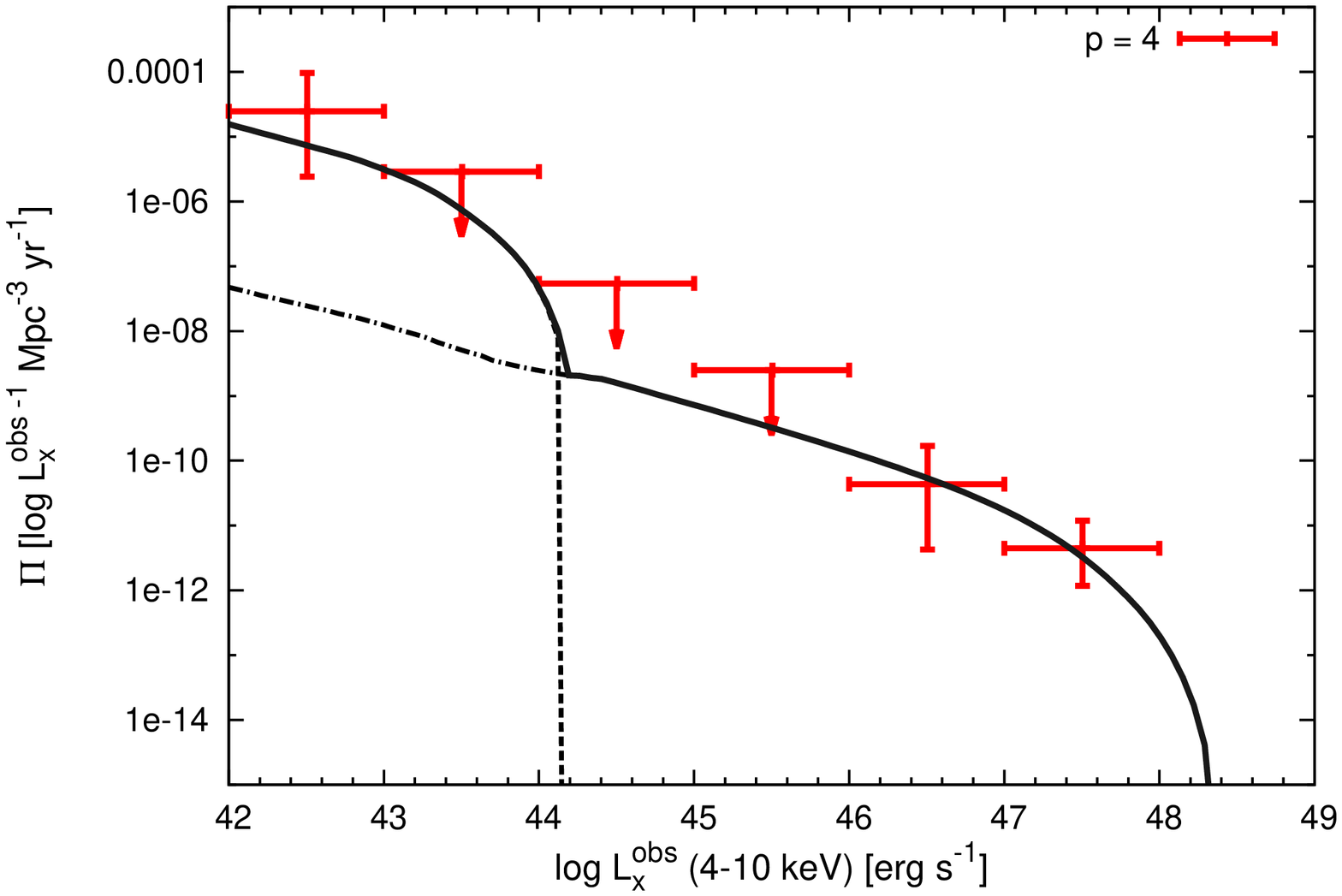}
\includegraphics[scale=0.35,angle=-90]{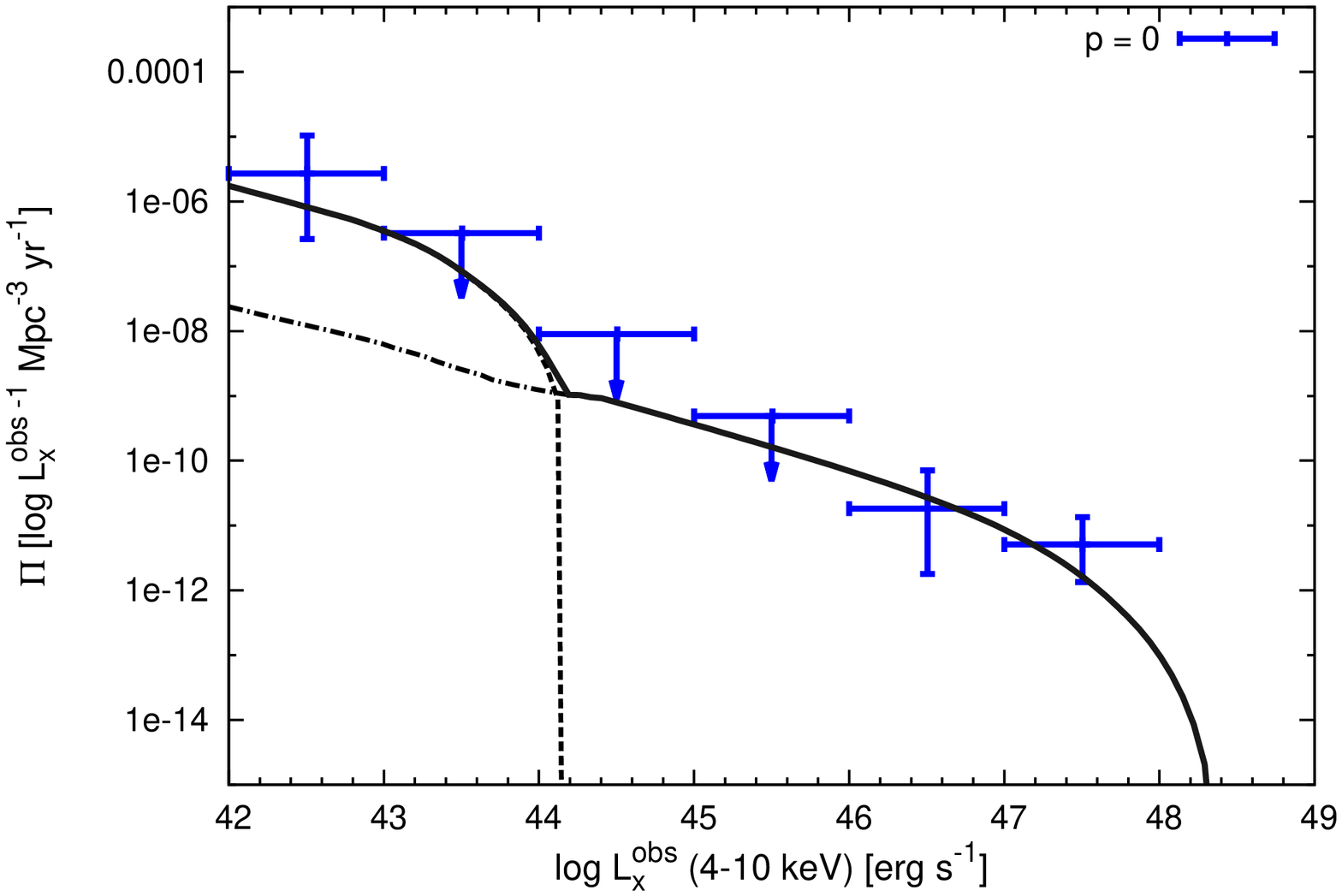}
\caption{
The best-fit XLF as a function of ``observed'' peak luminosity at $z=0.75$ 
for two evolution indices \textcolor{black}{(left figure: $p = 4$, right figure: $p=0$).} 
The solid lines represent the total XLF consisting of that of TDEs without jets (dotted line) 
and that of TDEs with jets (dot-dashed line).
}
\label{fig:ml_fit}
\end{figure}

Figure~\ref{fig:ml_fit} displays the results of the TDE XLF as a
function of $L^{\rm obs}_{\rm x}$ (observed peak luminosity in the 4--10 keV band), 
obtained for the cases of $p = 0$ and $4$.
The solid curves plot the best-fit model at $z=0.75$, which is obtained
by integrating ${\rm d}\Phi (L_{\rm x}, z=0.75)/{\rm d} L_{\rm x}$
over the half solid angle with respect to the jet direction, 
the mass of disrupted stars, 
and \lx\ (intrinsic peak luminosity of the Comptonized component) as 

\begin{footnotesize}
\begin{eqnarray}
&  \Pi (\log & L^{\rm obs}_{\rm x}, z=0.75)   \nonumber \\
& &  = \ln(10)~L^{\rm obs}_{\rm x} \int \int \int  {\rm d} L_{\rm x}  
{\rm d}\log M_\ast \frac{{\rm d}\Omega}{2\pi} \nonumber \\ 
& & \times \Bigl\{(1- f_{\rm jet})\delta_{\rm D} (C_0 L_{\rm x } - L^{\rm obs}_{\rm x})  
+ f_{\rm jet} \delta_{\rm D} 
\Bigl(C_1 L_{\rm x } - {L^{\rm obs}_{\rm x}} \Bigr) \Bigr\} \nonumber \\
& & \times \frac{{\rm d}\Phi (L_{\rm x}, z=0.75)}{{\rm d} L_{\rm x}} P(M_\ast).
\nonumber \\
\end{eqnarray}
\end{footnotesize}
The data points are plotted by the ``$N^{\rm obj}/N^{\rm model}$''
method \citep{Miya01}; they are calculated as 
\begin{footnotesize}
\begin{eqnarray}
\Pi (\log L^{\rm obs}_{\rm x}, z=0.75) \times \frac{N^{\rm data}}{N^{\rm model}},
\end{eqnarray}
\end{footnotesize}
where $N^{\rm data}$ is the number of observed events in each luminosity bin and
$N^{\rm model}$ is that predicted by a model. 
The error bars reflect the 90\% confidence level in $N^{\rm data}$ 
based on the formula of \citet{Ge86}. If no event is detected, 
we plot the 90\% upper limit by setting $N^{\rm data} = 2.3$ \citep{Ge86}.

Table~\ref{tab:tde_rate} lists the TDE occurrence rates per unit volume
(${\rm Mpc}^{-3}$ yr$^{-1}$) in different luminosity ranges predicted
from our best-fit TDE XLFs. The attached errors are calculated by only
considering the uncertainty in the normalization of the TDE XLF.

\begin{table}
\caption{Frequency of TDE occurrence at different luminosity ranges\label{tab:tde_rate}}
\begin{center}
\begin{tabular}{ccc}
\hline
\rule{-3pt}{3ex}  log\,{$L^{\rm obs}_{\rm x}$}  & $\dot{N}_{p = 0}$  & $\dot{N}_{p = 4}$  \\ 
  $[1]$ &  [2] & [3]   \\ \hline 
40-41 &         $1.2^{+1.2}_{-0.7}\times10^{-5}$ &      $1.1^{+1.1}_{-0.6}\times10^{-5}$ \\ \rule{-3pt}{2.3ex}
41-42 &         $4.0^{+4.0}_{-2.3}\times10^{-6}$ &      $3.8^{+3.8}_{-2.2}\times10^{-6}$ \\ \rule{-3pt}{2.3ex}
42-43 & $9.0^{+9.0}_{-5.1}\times10^{-7}$ &      $8.6^{+8.6}_{-4.8}\times10^{-7}$ \\ \rule{-3pt}{2.3ex}
43-44 & $1.2^{+1.2}_{-0.7}\times10^{-7}$ &      $1.1^{+1.1}_{-0.6}\times10^{-7}$ \\ \rule{-3pt}{2.3ex}
44-45 & $1.1^{+1.1}_{-0.6}\times10^{-9}$ &      $4.8^{+4.7}_{-2.7}\times10^{-10}$ \\ \rule{-3pt}{2.3ex}
45-46 & $1.8^{+1.8}_{-1.0}\times10^{-10}$ &     $3.8^{+3.8}_{-2.2}\times10^{-11}$ \\ \rule{-3pt}{2.3ex}
46-47 & $3.1^{+3.1}_{-1.8}\times10^{-11}$ &     $6.6^{+6.6}_{-3.7}\times10^{-12}$ \\ \rule{-3pt}{2.3ex}
47-48 & $2.6^{+2.5}_{-1.4}\times10^{-12}$ &     $5.4^{+5.4}_{-3.1}\times10^{-13}$ \\ 
\hline
\multicolumn{1}{@{}l@{}}{\hbox to 0pt{\parbox{160mm}
{\footnotesize
\textbf{Notes.}\\
Col. [1]: The luminosity range. \\
Col. [2]: The frequency of the TDE occurrence $({\rm Mpc}^{-3} {\rm yr}^{-1})$  \\
in the corresponding luminosity range for $\log L_{\rm x \ast} = 44.6$ and $p = 0$.\\
Col. [3]: The same as Col. [2] but for $p = 4$.\\
}\hss}}
\end{tabular}
\end{center}
\end{table}

\subsection{Comparison with {\it ROSAT} Results}\label{sec:comp_rosat}

We check the consistency of our results with a previous study based
on the {\it ROSAT} All-Sky Survey (RASS; \cite{Don02}). Here, we must
take into account the different TDE survey conditions between {\it MAXI}
and {\it ROSAT}. Because {\it MAXI} is continuously monitoring the entire sky, we can
derive the whole light curve of each TDE and hence its ``peak''
luminosity. By contrast, the detection of TDEs reported by the {\it ROSAT} survey was
based on two snapshot observations, the one in the scanning mode
during the RASS and the other in the pointing mode. Thus, in the
case of {\it ROSAT}, it is impossible to accurately estimate the ``peak''
luminosity of each TDE because of the uncertainty in its peak flux time
due to the scarce observations.

To compare our {\it MAXI} results with the {\it ROSAT} one, we need to
convert the XLF of TDEs given as a function of ``peak luminosity'' into
 an ``instantaneous'' XLF, which gives the probability of detecting
TDEs with an instantaneous luminosity {\it in a single epoch}.
Following \citet{Milo06}, for a given instantaneous luminosity of
$L'^{\rm obs}_{\rm x,ins}$, we formulate the instantaneous XLF as

\begin{scriptsize}
\begin{eqnarray}
&  \frac{{\rm d}\Phi'(L'^{\rm obs}_{\rm x,ins},z)}{{\rm d} \log L'^{\rm obs}_{\rm x,ins}} & = \ln (10)~L'^{\rm obs}_{\rm x,ins}
\int^{\Omega=2\pi}_{\Omega=0} \frac{{\rm d}\Omega}{2\pi} 
\int^{L_{\rm x, max}}_{L_{\rm x, min}} {\rm d}L_{\rm x} \int^\infty_{t_{\rm p}} {\rm d}t 
\int^{M_{\ast, \rm max}}_{M_{\ast, \rm min}}  {\rm d}\log M_\ast  \nonumber  \\
& & \times \frac{{\rm d} \Phi (L_{\rm x}, z)}{{\rm d} L_{\rm x}}  P(M_\ast) \nonumber \\ 
& & \times \Bigl\{(1- f_{\rm jet})  \delta_{\rm D} \Bigl(L'^{\rm obs}_{\rm x,ins} - L^{\rm obs}_{\rm x,ins} (C_0, L_{\rm x}, t) \Bigr) e^{- (1+z)^p (1 - f_{\rm jet})\xi(L_{\rm x})t}  \nonumber \\
& & \hspace{0.5cm} + f_{\rm jet} \delta_{\rm D} \Bigl(L'^{\rm obs}_{\rm x,ins} - L^{\rm obs}_{\rm x,ins} (C_1, L_{\rm x}, t) \Bigr) e^{- (1+z)^p~f_{\rm jet}~\xi(L_{\rm x})t}  \Bigr\}.
\label{eq:obs_lf}
\end{eqnarray}
\end{scriptsize}
The above equation takes into account the Poisson weighted probability 
along the luminosity decay. The peak time $t_{\rm p}$ is chosen to be 
0.1 yr, corresponding to the averaged value of our sample. 

To perform this calculation, we need $C_0$ and $C_1$, the conversion
factors from an intrinsic luminosity to an observed luminosity in the
the {\it ROSAT} band (0.2--2.4 keV). Since the blackbody component can
be dominant in this energy band, the term of $\omega_{\rm bb}$ in
equations (\ref{model_0}) and (\ref{model_1})  must be estimated. We 
follow our assumption that the rest of the non-beamed bolometric 
luminosity from which the Comptonization component is subtracted is 
dominated by a blackbody component with a single temperature. Accordingly, 
we estimate the effective temperature by assuming that the emitting 
area of the blackbody component is $\sim \pi (3 R_{\rm Sch})^2$.  
The color temperature is assumed to be identical to the effective one.  
As a result, $\omega_{\rm bb}$ is derived as a function of $L_{\rm x}$ 
and $z$. We obtain $\omega_{\rm pow} = 2.71$ for a power-law photon index of $2.0$. 


The predicted number of TDEs in a single epoch observation, $N_{\rm
TDE}$, is calculated as

\begin{scriptsize}
\begin{eqnarray}
N_{\rm TDE} = \int^{\log L'^{\rm obs}_{\rm x, max}}_{\log L'^{\rm obs}_{\rm x, min}} \int^{z_{\rm max}}_{z_{\rm min}}
\frac{{\rm d} \Phi' (L'^{\rm obs}_{\rm x,ins}, z)}{{\rm d} \log L'^{\rm obs}_{\rm x,ins}} \frac{{\rm d}V(
L'^{\rm obs}_{\rm x,ins}, z)}{{\rm d}z} {\rm d} z {\rm d} \log L'^{\rm obs}_{\rm x,ins}. \nonumber 
\label{n_tde}
\end{eqnarray}
\end{scriptsize}
Here, ${\rm d}V/{\rm d}z$ is the survey volume per unit redshift, which
is based on the analysis by \citet{Don02}. They detected five large-amplitude X-ray outbursts
by combining the RASS and pointed {\it ROSAT} observations. Their survey covered 
$\approx 9\%$ of the sky, which is complete to the flux limit of $2\times10^{-12}$ 
erg cm$^{-2}$ s$^{-1}$ (0.2--2.4 keV). If this flux limit is imposed, three (WPVS 007, 
IC 3599, and RX J1624.9+7554) out of the five TDEs are left to constitute a complete
sample. We note that recently \citet{Gru15} have reported that the event of IC 3599 
may not be a true TDE; in this case, the number of the {\it ROSAT} complete sample is 
reduced to two. In accordance with the study by \citet{Don02}, 
the threshold of the observed luminosity for the identification of TDEs is 
set to $\log L'^{\rm obs}_{\rm x, min} = 41$.

When comparing the {\it ROSAT} and {\it MAXI} results, we should take
into account the fact that the RASS performed in the soft X-ray band
would easily miss obscured TDEs, unlike the case of 
{\it MAXI}. Indeed, the follow-up observations of our four TDEs with {\it Swift}
or {\it XMM-Newton} indicate that two of them ({\it Swift} J1644+57 and
NGC 4845; \cite{Bur11}; \cite{Niko13}) are obscured with column
densities of $N_{\rm H} > 10^{22}$ cm$^{-2}$. From this result, we estimate the 
obscuration fraction to be $\sim 1/2$, and accordingly, decrease the detectable 
number of TDEs $N_{\rm TDE}$ by a factor of 2 to be compared with the {\it ROSAT} 
result.
\textcolor{black}{
Also, we implicitly assume that all TDEs have hard X-ray components. Although 
they are not significantly required in the RASS spectra of
WPVS 007 \citep{Gru95} and RX J1624.9+7554 \citep{Gru99}, their expected contribution
to the 0.5--2 keV flux can be very small ($\sim$1\%) and hence cannot
be well constrained with these data owing to the limited energy band
and photon statistics. To confirm this assumption, future broadband
observations of TDEs, like those by {\it eROSITA}, will be important.
}

We calculate $N_{\rm TDE}$ with equation (\ref{n_tde}) from our best-fit XLF models 
summarized in Table~\ref{best_fit}. We obtain $N_{\rm TDE}/2 = 3.2^{+3.2}_{-1.8}$ 
for $p = 0$ and $N_{\rm TDE}/2 = 4.0^{+3.9}_{-2.2}$ for $p = 4$, which are consistent 
with the observed number of TDEs (three or two) in the {\it ROSAT} survey. 
When $\lambda_{\rm Edd} = 0.1$ (corresponding to $\log L_{\rm X \ast} = 44.0$) is adopted 
instead of $\lambda_{\rm Edd} = 1.0$ ($\log L_{\rm x \ast} = 44.6$), the derived XLF 
predicts $N_{\rm TDE}/2 = 0.7^{+0.7}_{-0.4}$ for $p = 0$ and $N_{\rm TDE}/2 = 1.2^{+1.2}_{-0.7}$ 
for $p = 4$. These numbers are significantly smaller than three at $90$\% confidence 
level, although that for $p=4$ is consistent with two (i.e., when the event in IC 3599 
is excluded from the {\it ROSAT} sample). 
\textcolor{black}{
Also, under the assumption of $\lambda_{\rm Edd} = 5$ and $k_{\rm 2-10} = 70$ 
as a super-Eddington accretion case, the predicted TDE number is significantly 
higher ($N_{\rm TDE}/2 \gtrsim 4$) than the observed one 
regardless of the evolution index ($p$). 
}
Hence, we adopt $\lambda_{\rm Edd} = 1.0$ in 
our baseline model, which is allowed for the range of $p=0$--$4$.

In the above calculations of $N_{\rm TDE}$, we have ignored possible
time evolution of the X-ray spectrum of each TDE during the decay
phase. According to \citet{Vas07}, the bolometric correction factor from
the 2--10 keV band ($k_{\rm 2-10}$) depends on Eddington ratio. Since 
$k_{\rm 2-10}$ determines the relative weights between the Comptonized
and blackbody components (see fourth paragraph in this Section) 
in our assumption, the X-ray spectrum is predicted to be time dependent. To roughly examine these
effects, we calculate $N_{\rm TDE}$ by approximating that $k_{\rm 2-10}
= 50$ and $k_{\rm 2-10} = 20$ when $\lambda_{\rm Edd} \geq 0.1$ and
$\lambda_{\rm Edd} < 0.1$, respectively. In the case of $\lambda_{\rm
Edd} = 1.0$, we obtain $N_{\rm TDE}/2 = 3.7^{+3.7}_{-2.1}$ for $p = 0$
and $N_{\rm TDE}/2 = 4.8^{+4.8}_{-2.7}$ for $p = 4$, while $N_{\rm
TDE}/2 < 3$ is obtained for both $p = 0$ and $4$ in the case of
$\lambda_{\rm Edd} = 0.1$. Hence, the possible spectral evolution does
not affect our conclusion.

\subsection{Effects by Changing Fixed Parameters}\label{sec:eff_par}

In this subsection, we examine the effects on the XLF results, in
particular $f_{\rm jet}$, by changing the fixed parameters in the XLF
model that are difficult to be constrained from the data: 
(1) the Lorentz factor of the jets ($\Gamma$), 
(2) the dependence of the specific TDE rate on SMBH mass ($\lambda$), 
(3) the fraction of the intrinsic jet luminosity in \lx\ ($\eta_{\rm jet}$), and 
(4) the upper mass boundary of tidally disrupted stars ($M_{\ast, {\rm max}}$). 
Considering the small sample size, we perform a ML fit of the XLF 
by adopting an alternative value instead of the default value 
for each fixed parameter and check how $f_{\rm jet}$ is affected. We 
consider $\Gamma =$ 5 and 20 (default is $10$), 
$\lambda = -0.1$ (default is $-0.4$), 
$\eta_{\rm jet} = 0.01, 1.0$ and $10.0$  (default is $0.1$), and 
$M_{\ast,{\rm max}}/M_\odot = 10$ and $100$ (default is $1$).  
In each case, we assume the two evolution indices, $p=0$ and $p=4$.
We regard that the minimum or maximum values considered here 
correspond to extreme cases within physically plausible values.

\begin{figure*}[htb]
\hspace{0.5cm}
\includegraphics[scale=0.25,angle=-90]{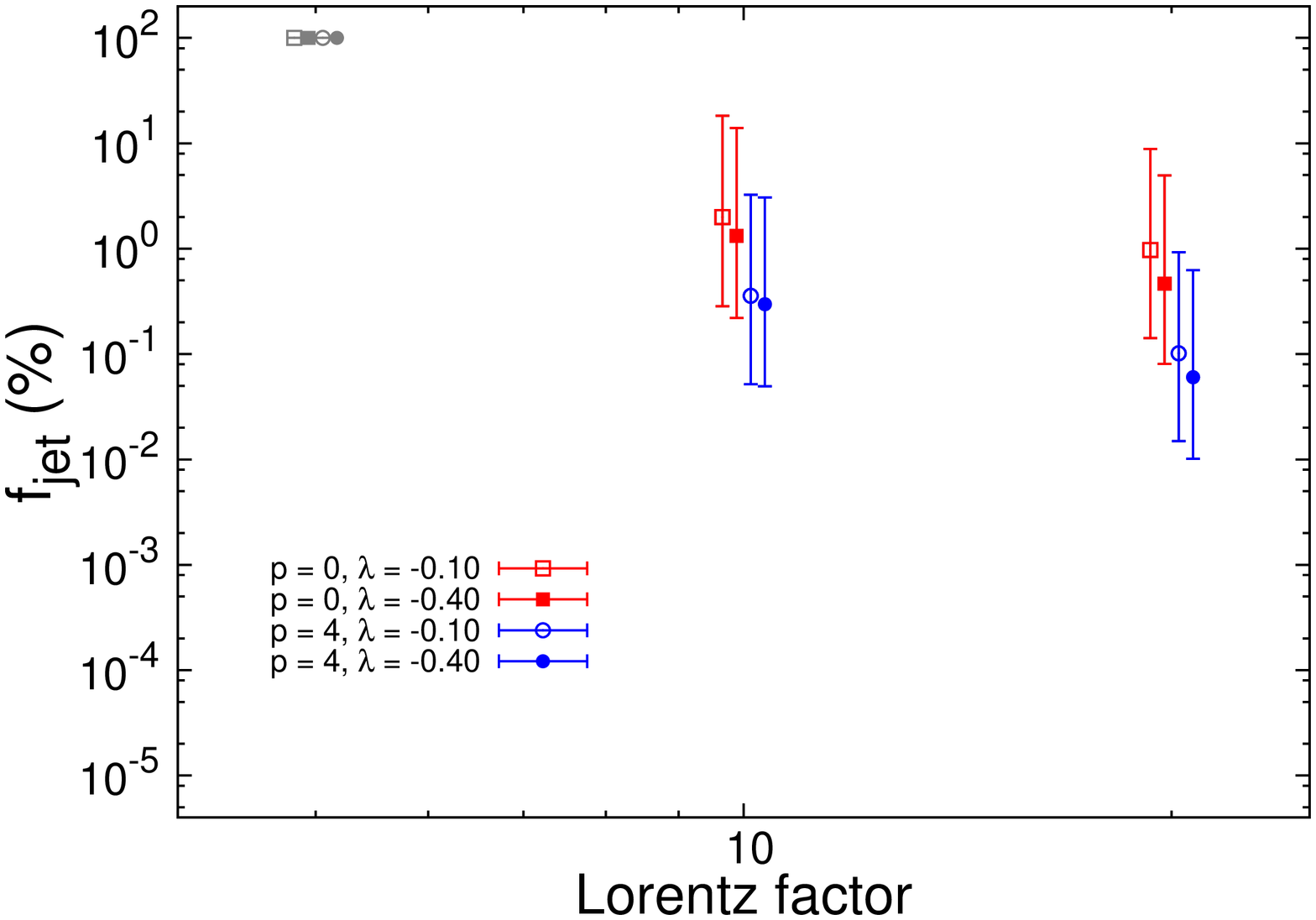}
\includegraphics[scale=0.25,angle=-90]{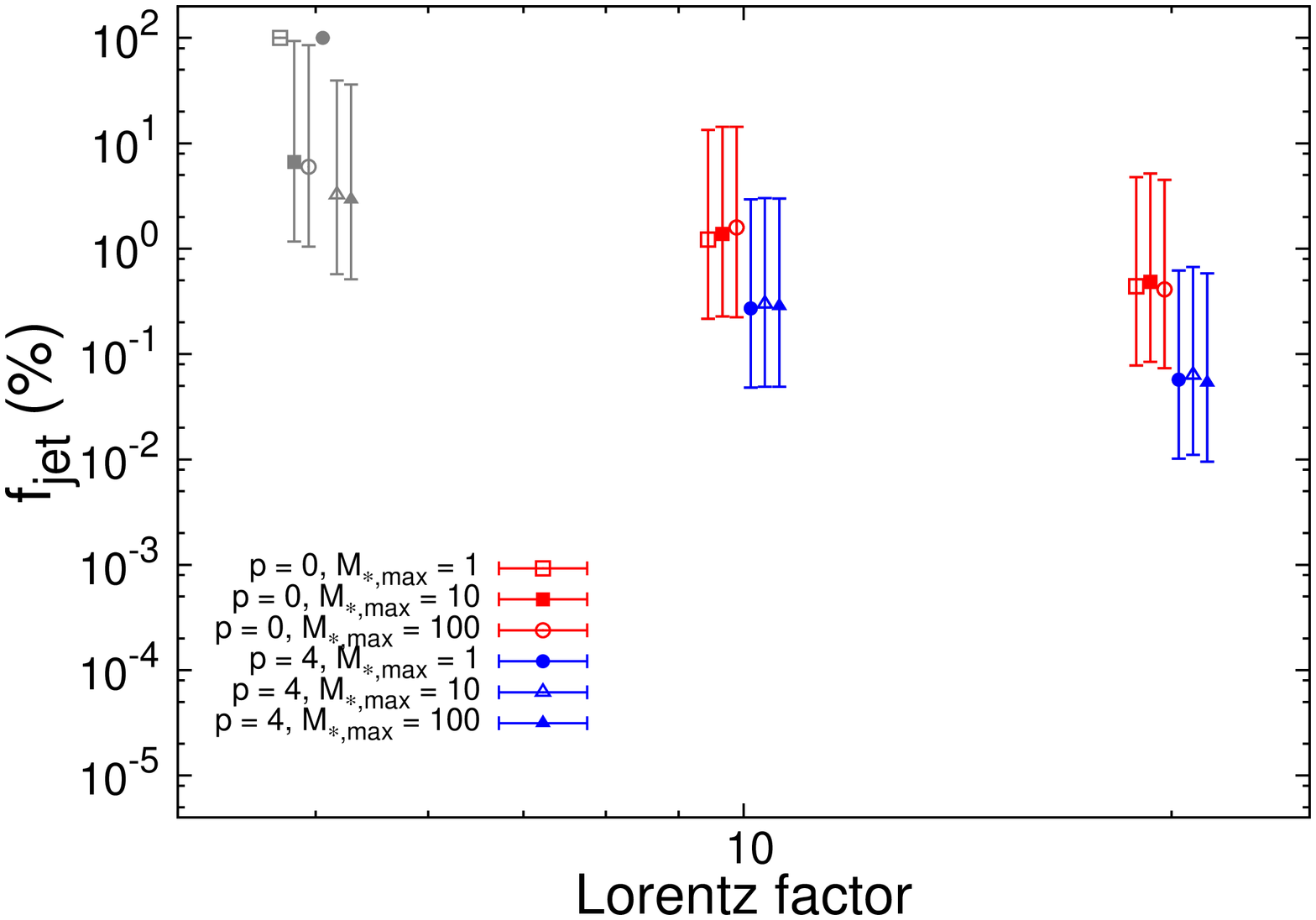}
\includegraphics[scale=0.25,angle=-90]{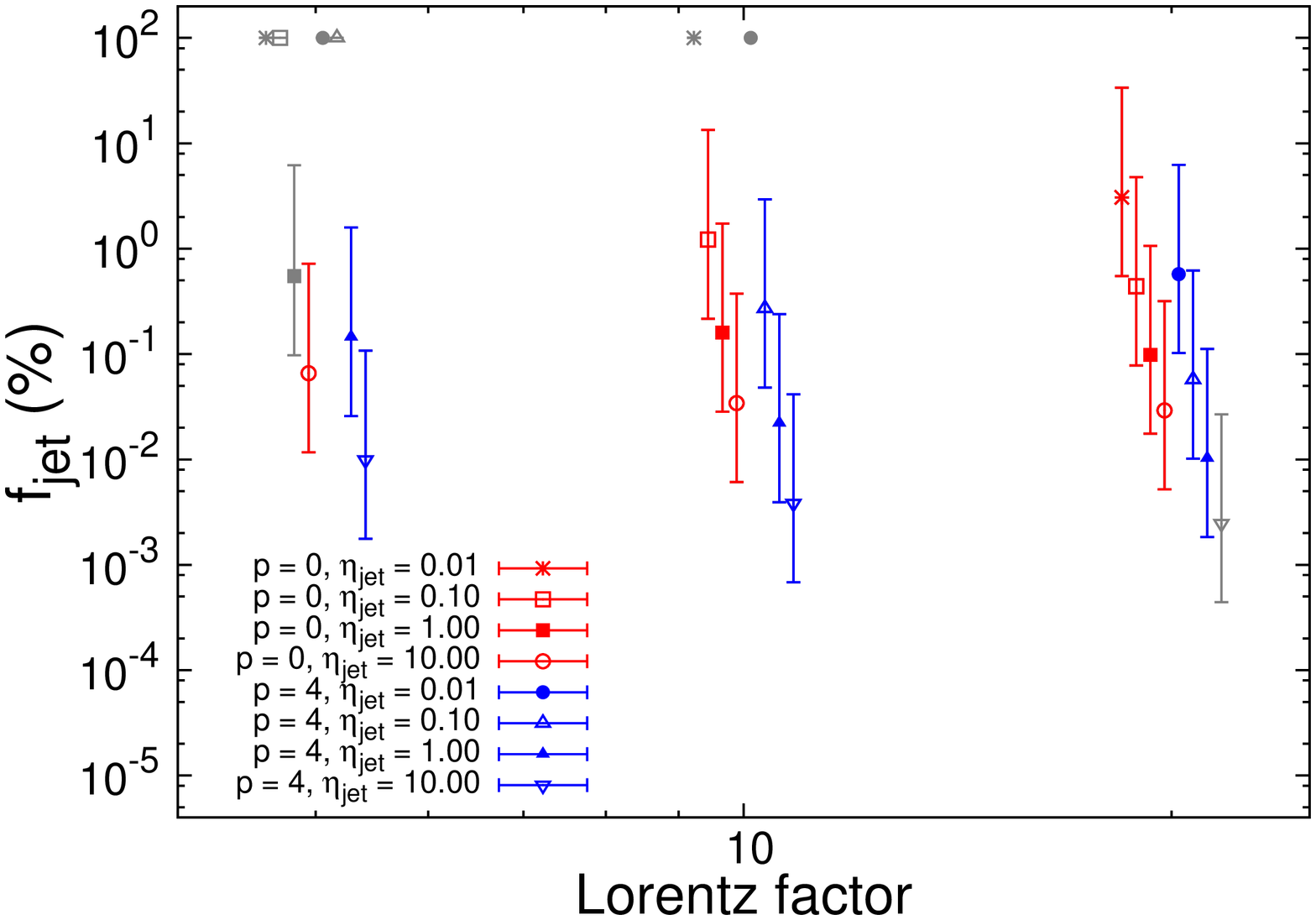} \vspace{0.5cm}
\caption{
The constraint on $f_{\rm jet}$ plotted against the jet Lorentz factor
$\Gamma$ for various choices of the other parameters in the XLF 
 ($\lambda$, $\eta_{\rm jet}$, $M_{\ast, {\rm max}}$). For clarity, the 
data points are slightly shifted along the $x$ axis. 
The gray points correspond to the case where the fit is rejected in
terms of the one-dimensional KS tests (for the redshift and/or
luminosity distributions), and/or the predicted number of TDEs in the
{\it ROSAT} survey. Also, we plot the points at the position of $f_{\rm
jet} = 100\%$ in gray when the TDE XLF cannot reproduce
the observed maximum luminosity ($\log L_{\rm X} = 47.5$). Hence, these
gray points should be ignored. } \label{fig:f_jet}
\end{figure*}

Figure~\ref{fig:f_jet} summarizes the constraints on $f_{\rm jet}$
plotted against $\Gamma$ when one of the other parameters ($\lambda$,
$\eta_{\rm jet}$, $M_{\ast, {\rm max}}$) is changed. 
The red and blue marks correspond to $p=0$ and $p=4$, respectively.
If an acceptable
fit is not obtained in terms of the one-dimensional KS tests (for the
redshift and luminosity distributions) and/or the predicted number of
TDEs in the {\it ROSAT} survey, we mark them in gray color. Also, when
the maximum luminosity in the TDE XLF is lower than the observed ones
($\log L_{\rm X}= 47.5$), we plot the points at the position of $f_{\rm jet} =
100\%$ in gray color. These gray points should be ignored because the
corresponding XLF model is rejected.

Among the acceptable parameter sets, we obtain the {\it maximum} upper-limit
of $f_{\rm jet}=34\%$ for $\Gamma = 20$, $\eta_{\rm jet} = 0.01$, and $p
= 0$, and the {\it minimum} lower-limit of $f_{\rm jet} = 0.0007\%$ for
$\Gamma = 10$, $\eta_{\rm jet} = 10.0$, and $p = 4$. Thus, we
conservatively estimate that the fraction of TDEs with jets among all
TDEs is $0.0007$--$34\%$. This constraint is compatible with the
fraction of radio loud AGNs in all AGNs ($\sim$10\%). We note that
$f_{\rm jet}$ depends on a combination of $\eta_{\rm jet}$ and
$\Gamma$. This is mainly because the cutoff luminosity of the XLF for
TDEs with jets is determined as $\eta_{\rm jet} \delta (\Gamma,\theta)^4$, 
with which $f_{\rm jet}$ is strongly coupled: one obtains a
large value of $f_{\rm jet}$ when $\eta_{\rm jet}$ and/or $\Gamma$
becomes smaller. Since $\Gamma$ determines the solid-angle of the
detectable relativistic jets ($\propto 1/\Gamma^2$), there is another
coupling of $f_{\rm jet}$ with $\Gamma$, which partially cancels out the
coupling through the cutoff luminosity.





\section{Discussions}\label{dis} 

\subsection{Results Summary}

Utilizing a complete sample of TDEs detected in the {\it MAXI}
extragalactic survey in the 4--10 keV band, we have, for the first
time, quantitatively derived the shape of the XLF of TDEs (the
occurrence rate of a TDE per unit volume) as a function of intrinsic 
peak luminosity at $z \lesssim 1.5$. Our XLF takes account of two TDE types, one with jets 
and the other without jets, and those subject to heavy
absorption that would be difficult to detect in soft X-ray
surveys. In the modelling of the XLF, we have taken into account the
mass function of SMBHs, that of disrupted stars, the specific TDE rate 
as a function of SMBH mass, and relativistic beaming effects from jets,
although we need to fix several parameters at reasonable values. The XLF
can be converted to an ``instantaneous XLF'' obtained from a single-epoch
observation in different energy bands, by assuming spectra of TDEs.
Our baseline model, whose parameters are listed in Table~\ref{tab:def_par}, is found
to well reproduce the number of TDEs previously detected in the {\it
ROSAT} survey. The main finding is 
that the fraction of TDEs with jets among all TDEs is  $0.0007$--$34\%$. 
Our result will serve as a reference model of a TDE XLF, which would be
useful to estimate their contribution to the growth history of SMBHs
and to predict TDE detections in future missions.

\subsection{Comparison of TDE Rate with Previous Studies}

We compare the TDE rate per unit volume based on our best-fit XLF with
those estimated from previous studies. Combining the {\it XMM-Newton}
slew survey and the RASS, \citet{Esq08} detected two TDEs, whose soft
X-ray luminosities were $\sim 5\times10^{41}$ erg s$^{-1}$ and $\sim
5\times10^{43}$ erg s$^{-1}$. They derived a TDE rate per unit volume to
be $\sim 5\times10^{-6}$ Mpc$^{-3}$ yr$^{-1}$ for the peak luminosity of 
$10^{44}$ erg s$^{-1}$ in the 0.2-2.0 keV band.  Our result in Table 
\ref{tab:tde_rate} predicts the rate of unobscured TDEs with the 
peak luminosity $>10^{41}$ erg s$^{-1}$ (4--10 keV) to be $\sim
5\times10^{-6}$ Mpc$^{-3}$ yr$^{-1}$ at $z = 0$, which is in good
agreement with the estimate by \citet{Esq08}.

\citet{Mak10} studied a TDE in the galaxy cluster Abell 1689, using 
{\it Chandra} and {\it XMM-Newton}. The TDE rate ``per galaxy'' was 
estimated to be $6\times 10^{-5}$ galaxy$^{-1}$ yr$^{-1}$ for the minimum 
luminosity in the 0.3--2.5 keV band of $10^{42}$ erg s$^{-1}$. Our XLF 
predicts a TDE rate of $\sim 1\times 10^{-4}$ galaxy$^{-1}$ yr$^{-1}$ for 
peak luminosities of $\log L_{\rm x} > 42$, by adopting the spatial density 
of galaxies of $\phi_0 = 0.007$ galaxy Mpc$^{-3}$. The discrepancy by a 
factor of $\sim$2 may be explained if only half number of member galaxies
in Abell 1689 can produce TDEs detectable by their analysis, as mentioned 
in \citet{Mak10}.

It is interesting to compare our TDE rate with that
obtained from the flux-complete, optically-selected sample by
\citet{Van14} using archival Sloan Digital Sky Survey data. They
detected two TDE candidates, and estimated the TDE rate of $\sim
(4-8)\times10^{-8}$ yr$^{-1}$ Mpc$^{-3}$ for SMBH masses of $\sim
10^{7}$ $M_\odot$. Our estimate for the corresponding luminosity ($\log
L_{\rm X} \gtrsim 43$) is $\sim 1\times10^{-7}$ yr$^{-1}$
Mpc$^{-3}$. This is higher than the TDE rate of \citet{Van14}. It may be
owing to dust extinction, which reduces the number of TDE flares
detectable in the optical band. Indeed, if we correct the optical TDE
rate for obscuration by a factor of $2$, it becomes $\sim
(8-16)\times10^{-8}$ yr$^{-1}$ Mpc$^{-3}$, which is consistent with the
X-ray result.

\subsection{Contribution of TDEs to X-ray Luminosity Functions of Active
  Galactic Nuclei}\label{cont_XLF}

\begin{figure*}[htb]
\includegraphics[scale=0.8]{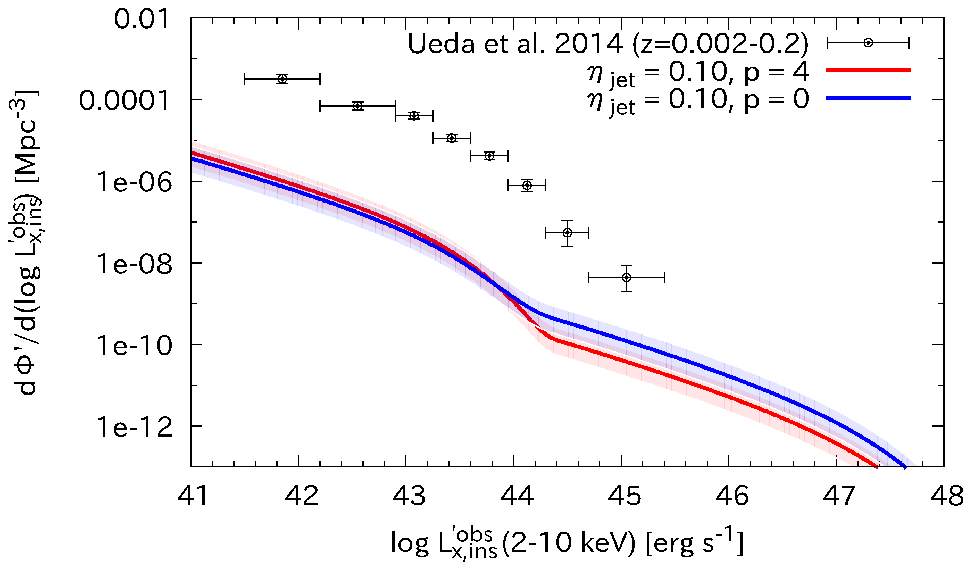} \hspace{0.5cm}
\includegraphics[scale=0.8]{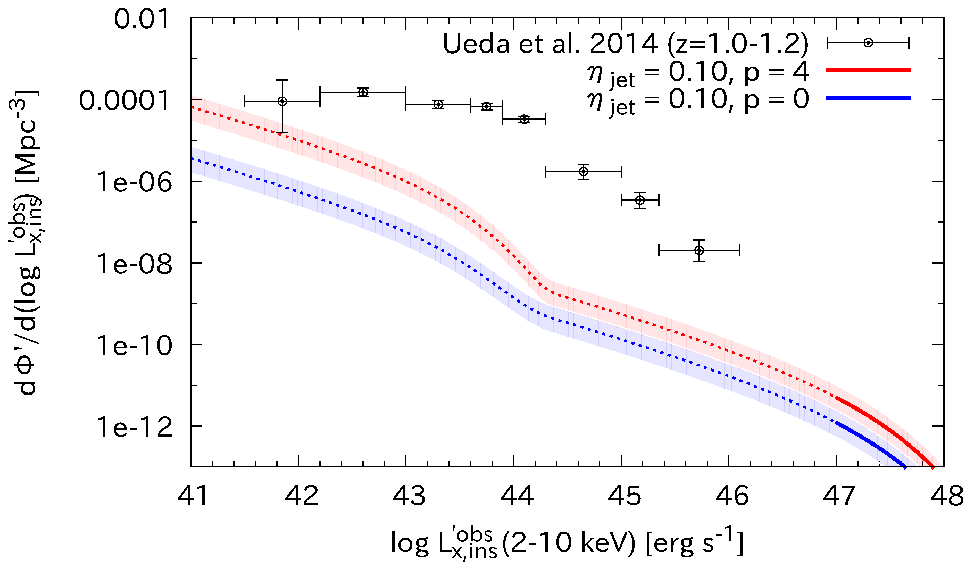} 
\caption{
Comparison of the instantaneous XLF of TDEs (our work) and the XLF of
AGNs \citep{Ueda14} in the 2--10 keV band. The upper and lower panels
show the results in the low $(z = 0.1)$ and high $(z = 1.1)$ redshift
ranges, respectively. The 1$\sigma$ Poisson errors are attached to the
AGN XLF.
}
\label{fig:obs_lf_hard}
\end{figure*}

It is possible that the XLF of AGNs may be
contaminated by TDEs, which are difficult to be distinguished from AGNs
in observations of limited numbers. To investigate this effect, we
calculate the instantaneous XLFs of TDEs based on our best-fit
parameters with equation (\ref{eq:obs_lf}). Figure~\ref{fig:obs_lf_hard}
plots the results in the 2--10 keV band at the low redshift ($0.002 < z
< 0.2)$ and high redshift $(1.0 < z < 1.2)$ ranges for the two evolution
indices, $p=0$ and $p=4$. The error region at 90\% confidence level due
to the uncertainty in the normalization is also indicated. The
luminosity range below the sensitivity limit where the TDE XLF is not
directly constrained from the {\it MAXI} survey is indicated by the
dashed lines.

For comparison, we overplot the hard XLF of AGNs obtained by 
\citet{Ueda14} at $z=0.1$ and $z=1.1$ for the two redshift ranges,
respectively. As noticed from Figure~\ref{fig:obs_lf_hard}, the
instantaneous XLF of TDEs are far below the observed AGN XLF at $0.002 <
z < 0.2$, indicating the contribution of TDEs is negligible in the local
universe. At $1.0 < z < 1.2$, the TDE contribution to the AGN XLF is
also negligible at high luminosities, while it could be significant at
the lowest luminosity range. However, the model of the TDE XLF in this
region is just an extrapolation from the result at the low redshift
range, and must be constrained by more sensitive surveys to reach any
conclusions.

\subsection{Mass Accretion History of SMBHs by TDEs}\label{mass_his}

TDEs contribute to the growth of SMBHs as argued by \citet{Sol82} for
AGNs. We here calculate the evolution of the mass density of SMBHs by TDEs
as done for AGNs (e.g., \cite{Mar04};
\cite{Sha04}). The bolometric luminosity of a TDE 
can be related to the mass accretion rate $\dot{M}_{\rm acc}$ via the 
mass-to-radiation conversion efficiency
$\epsilon$ as
\begin{equation}
L_{\rm bol} = \epsilon c^2 \dot{M}_{\rm acc}.
\end{equation}
The mass growth rate of the SMBH is given by
\begin{equation}
\dot{M}_{\rm BH} = (1 - \epsilon) \dot{M}_{\rm acc}.
\end{equation}
We formulate the SMBH mass-density equation as 
\begin{scriptsize}
\begin{eqnarray}\label{mass_dens}
& \rho (z) = 
\int^{z}_{z_{\rm s}} {\rm d}z \frac{{\rm d}t}{{\rm d}z} 
\int^{L_{\rm x, max}}_{L_{\rm x,  min}} {\rm d} 
L_{\rm x} \frac{{\rm d} \Phi (L_{\rm x}, z)}{{\rm d} L_{\rm x}}   
\int^{M_{\ast, \rm max}}_{M_{\ast, \rm min}} P(M_\ast) {\rm d}\log M_\ast\nonumber \\
& \times \frac{1-\epsilon}{\epsilon c^2}  \int^{\infty}_{t_{\rm p}} 
\{ (1-f_{\rm jet}) L_{\rm peak} + f_{\rm jet} L^{\rm jet}_{\rm peak} \} \Big( \frac{t}{t_{\rm p}}\Big)^{-5/3} {\rm d}t,
\end{eqnarray} 
\end{scriptsize}
where $z_{\rm s}$, $L_{\rm peak}$, and $L^{\rm jet}_{\rm peak}$ are the 
initial redshift from which the calculation starts, 
the peak bolometric luminosity without jets, and that with jets. 
Here, $L_{\rm peak}$ is set to the Eddington luminosity, while $L^{\rm jet}_{\rm peak}$ is 
the sum of the Eddington luminosity and intrinsic jet luminosity. In equation (\ref{mass_dens}), 
we assume that 
the time taken for most of the mass of a disrupted star to fall onto the SMBH is shorter 
than the cosmological time scale. 

\begin{figure}[!ht]
\hspace{0.5cm}
\includegraphics[scale=0.35,angle=-90]{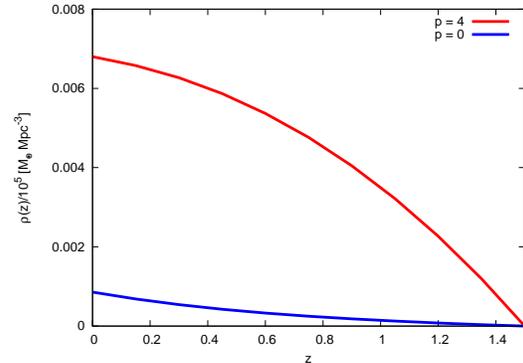}
\caption{
The evolution of SMBH mass-density caused by TDEs. 
Each line is colored in accordance with the notation in the figure.
}
\label{mass_hist}
\end{figure}

We adopt the 
mass-to-radiation conversion efficiency
of $\epsilon = 0.1$
similarly to the case of AGNs. Figure~\ref{mass_hist} shows the cumulative
SMBH mass-density $(M_\odot$ ${\rm Mpc}^{-3})$ calculated from $z_{\rm s}=1.5$. We find
that the total mass density at $z=0$ is at most $ 7\times10^2 M_\odot$
Mpc$^{-3}$ even for the case of $p =$ 4. This indicates that the SMBH
mass-density contributed by TDEs is much less than that of AGNs (e.g.,
\cite{Ueda14}). This is what is expected from the comparison of
the XLF between TDE and AGN as described in the previous subsection.

\section{Summary}\label{sum} 

We have derived the XLF of TDEs, i.e., the occurrence rate of a TDE per unit
volume as a function of intrinsic peak luminosity, from the {\it MAXI}
extragalactic survey. Our sample consists of four TDEs, {\it Swift} J1112-82, {\it Swift}
J1644+57, {\it Swift} J2058+05, and NGC 4845, detected in the first 37
months data of {\it MAXI}/GSC at high Galactic latitudes $(|b|
>10^\circ)$. It is complete to a flux limit of $\sim 3\times 10^{-11}$
erg s$^{-1}$ cm$^{-2}$ (4--10 keV), and is less affected with 
absorption than those detected in lower energy bands. In fact, two out
of the four TDEs show significant absorptions in the X-ray spectra.

We formulate the shape of the TDE XLF, based on the mass function of
SMBHs, that of tidally disrupted stars, and the specific TDE rate as a 
function of SMBH mass. We take account of two distinct types of TDEs, 
those with jets and those without them, and also the relativistic beaming 
from the jets. To incorporate effects of the cosmological evolution, 
we assume two cases where the XLF is constant over redshift or is
proportional to $(1+z)^4$. ML fits are performed to the observed TDE
sample, with the normalization of the XLF (i.e., TDE rate) and the
fraction of TDEs with jets among all TDEs, $f_{\rm jet}$, allowed to vary.
We then verify the best-fit model by checking consistency with the 
{\it ROSAT} study by \citet{Don02}. Consequently, we find 
that $f_{\rm jet}$ is constrained to be $0.0007$--$34$\%, consistent with
the case of AGNs.

On the basis of our best-fit TDE XLF, we have estimated the contribution
of TDEs to instantaneous XLFs of AGNs and to the evolution of the SMBH
mass density. It is found to be much smaller than those of AGNs, indicating 
that the effect by TDEs to the growth of SMBHs is negligible at $z \lesssim 1.5$. 
Future observations of TDEs, including the {\it eROSITA} survey, will enable 
us to establish more accurate statistical properties of TDEs over wide 
luminosity and redshift ranges.

\bigskip 

Part of this work was financially supported by the Grant-in-Aid for JSPS
Fellows for young researchers (TK) and for Scientific Research 26400228
(YU). This research has made use of {\it MAXI} data provided by RIKEN, JAXA 
and the {\it MAXI} team.


\appendix
\setcounter{table}{0}
\setcounter{figure}{0}
\def\thesection{\Alph{section}}
\def\thetable{\Alph{table}}

\section{Detection of X-ray Transient Events from {\it MAXI} Data}\label{app:sec:ta_search}

This section describes the image analysis of the {\it MAXI}/GSC data employed
to extract the transient X-ray events. The method is essentially the
same as that of \citet{Hiroi13} used to produce the 37-month catalog, but
is applied to 37 (or 12) individual images binned in 30 days (or 90
days) to obtain the light curves of all sources including newly detected
transient objects. 

The entire all-sky image in each time bin is divided into 768 areas of
14$^\circ \times$14$^\circ$ size centered at the coordinates defined in
the HEALPix software \citep{Gor05}. Circular regions around the very
bright sources Sco X-1, Cyg X-2, and Crab Nebula are not used in our
analysis to avoid systematic errors in the calibration of the point
spread function (PSF). First, combining the observed image and the model
of the background (the cosmic X-ray background plus the non X-ray
background), we make the significance map; an example is shown in
Figure~\ref{detection_ex}. Here the significance at each position is
simply calculated as the background-subtracted counts divided by the
square root of total counts in the 0$^\circ$.1$\times$0$^\circ$.1 region
around it. Then, excess points with the peak significance above
5.5$\sigma$ are left as source candidates. We regard those whose
positions do not match any sources in the 37-month {\it MAXI}/GSC catalog
within 1$^\circ$ as candidates of transient sources newly detected in
this time-sliced image analysis.

To determine the fluxes of all source candidates, we then perform image 
fitting by a model composed of the background and PSFs. Here we consider 
PSFs of all sources in the 37-month {\it MAXI} catalog and those of the 
transient candidates extracted above. The PSFs are calculated with the 
{\it MAXI} simulator \citep{Egu09} by assuming the spectrum of the Crab 
nebula. The fluxes of all sources (in units of Crab) and the normalization 
of the background  are left as free parameters.  The positions of 
the 37-month {\it MAXI} catalog sources are fixed according to the
results by \citet{Hiroi13}, while those of the transient candidates are
set to be free parameters.  To find the best-fit parameters and their
statistical errors, we employ the maximum likelihood algorithm based on
the C statistics \citep{Cash79}, utilizing the MINUIT software package.
To ensure complete detections of transient events, we repeat the above
procedures twice for each image. Namely, we again make the significance
map based on the best-fit model including the PSFs, search for remaining
residuals with the significance above 5.5$\sigma$, and then perform the
image fitting by including new source candidates.

As the results, we detect 10 transient events in total at high galactic
latitudes ($|b| > 10^\circ$), whose detection significance exceeds 5.5
in either of the analyzed images. Here we have paid careful attention to
exclude fake events caused by the Sun light contamination by checking
the image and spectra. Table \ref{all_te_info} summarizes the basic
information of the events. The parameters are derived from an image
where the event was detected with the highest detection significance
among all time bins.

\begin{figure*}[!ht]
\includegraphics[scale=0.27,angle=-90]{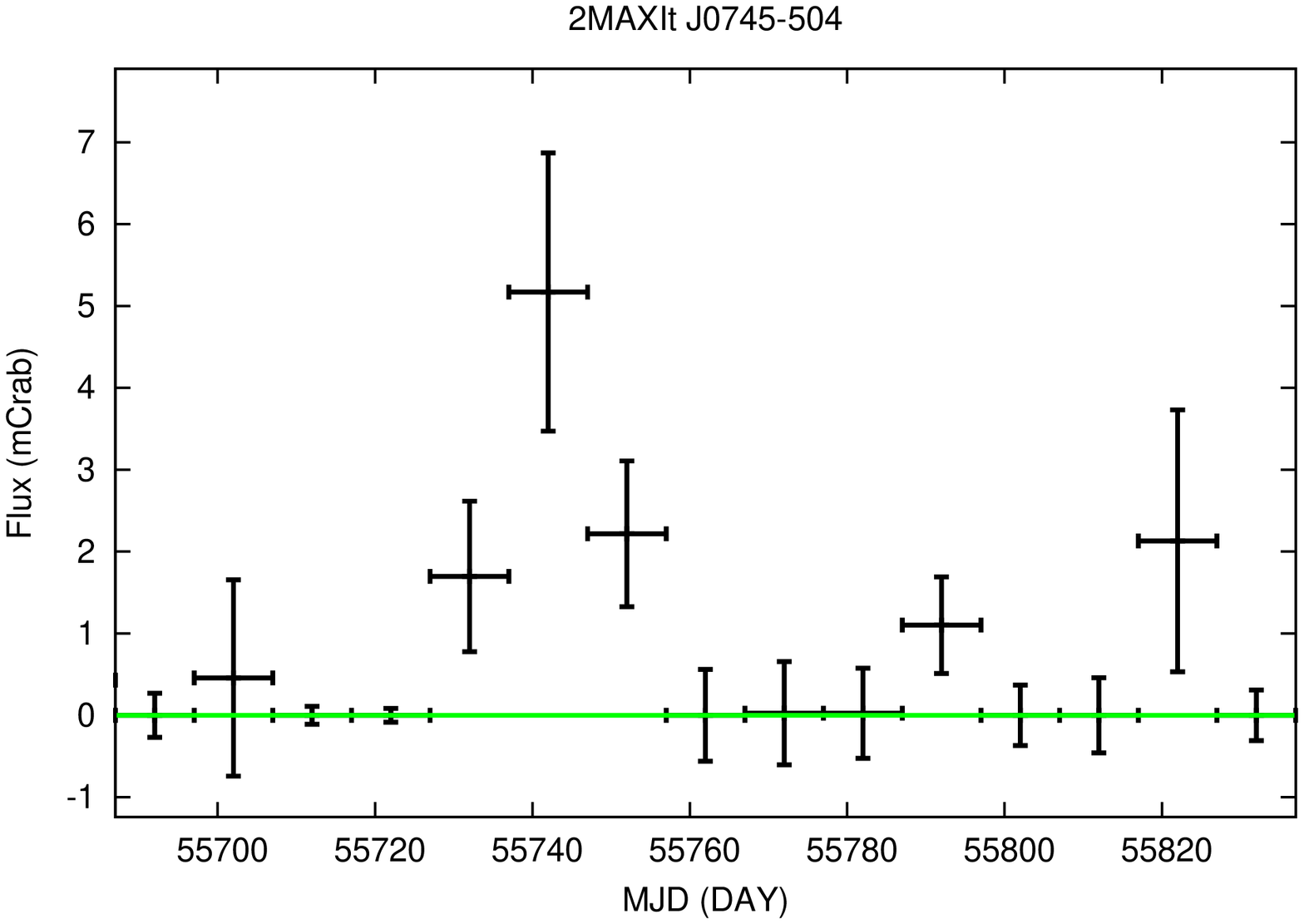}
\includegraphics[scale=0.27,angle=-90]{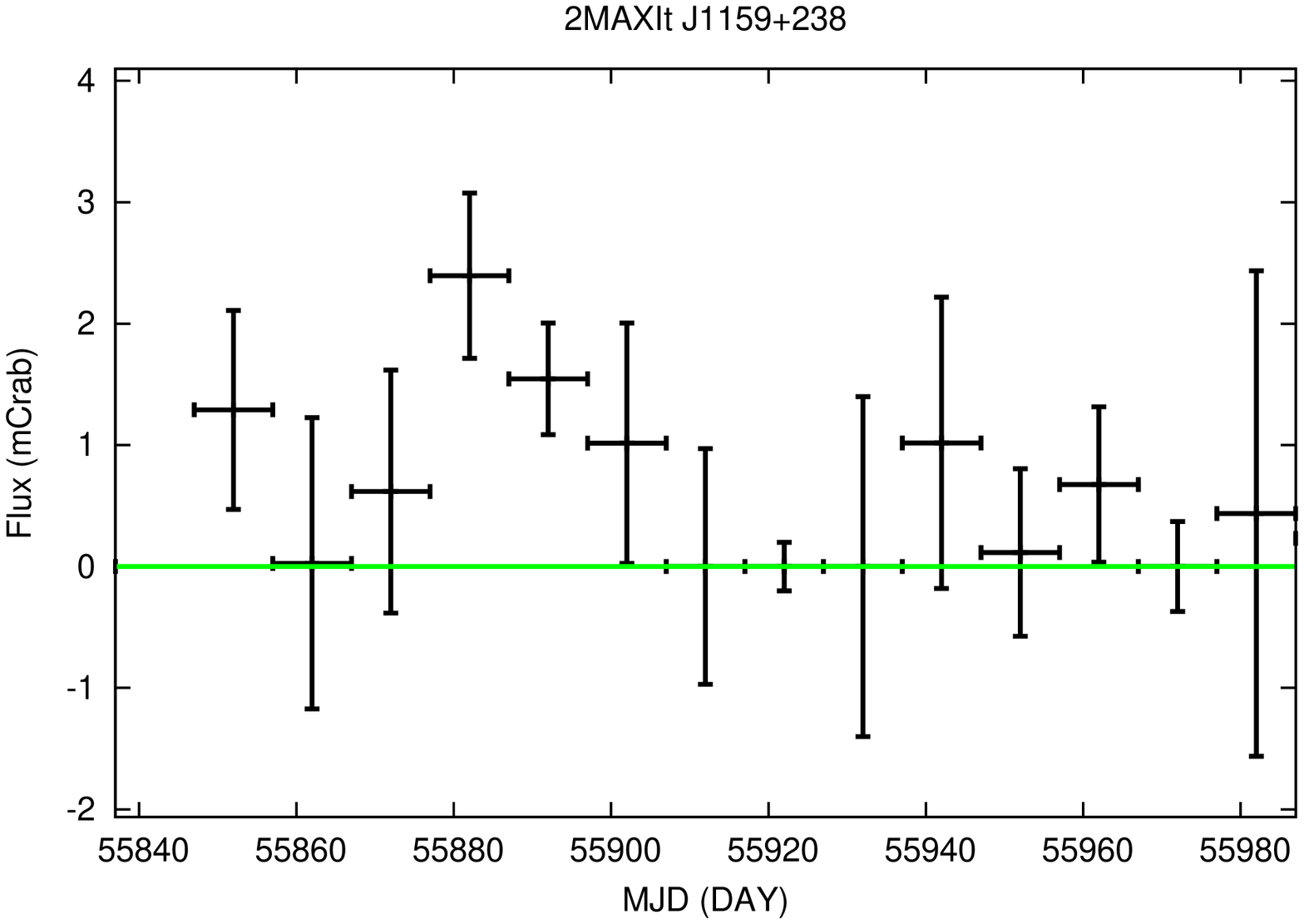}
\includegraphics[scale=0.27,angle=-90]{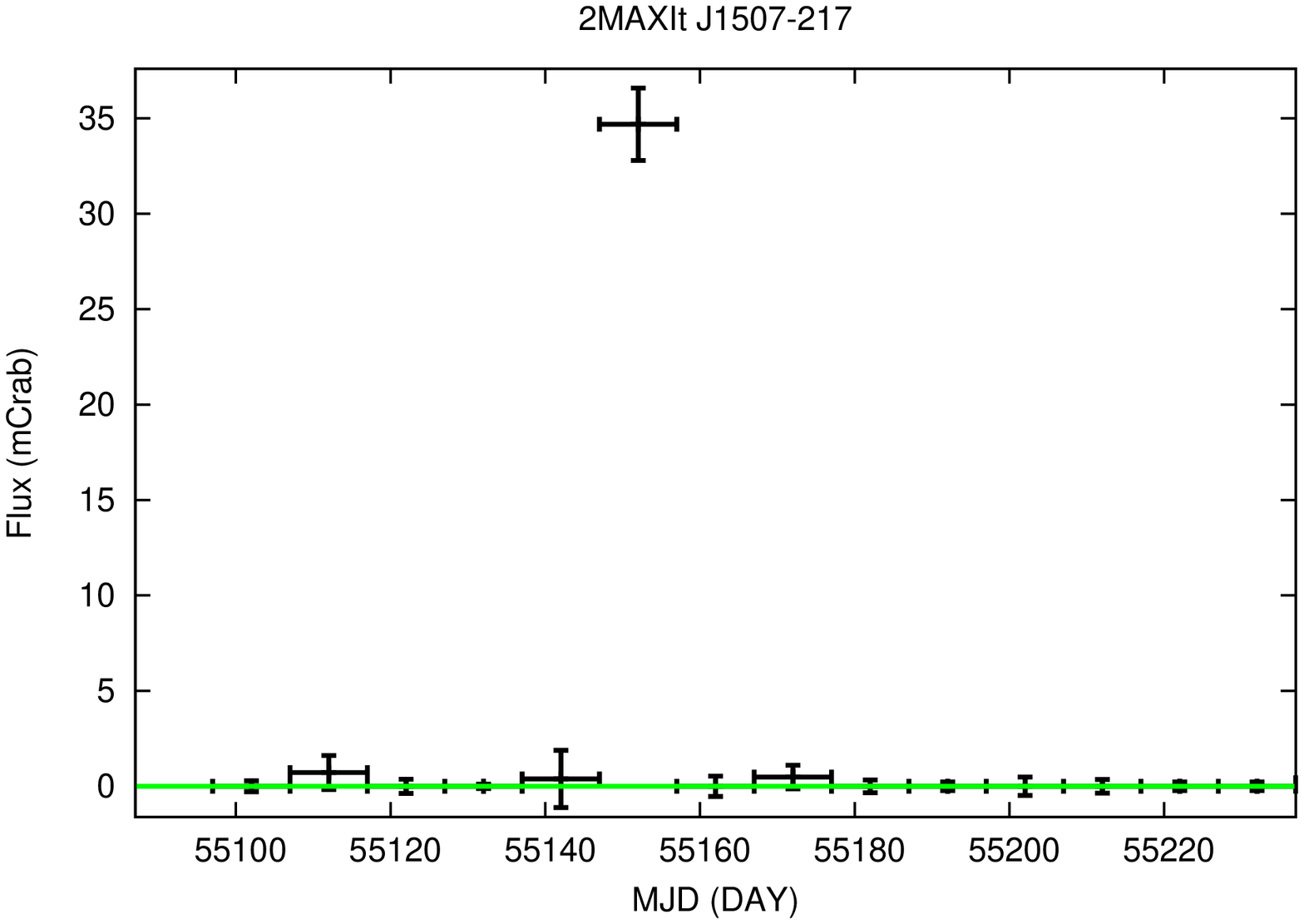}
\includegraphics[scale=0.27,angle=-90]{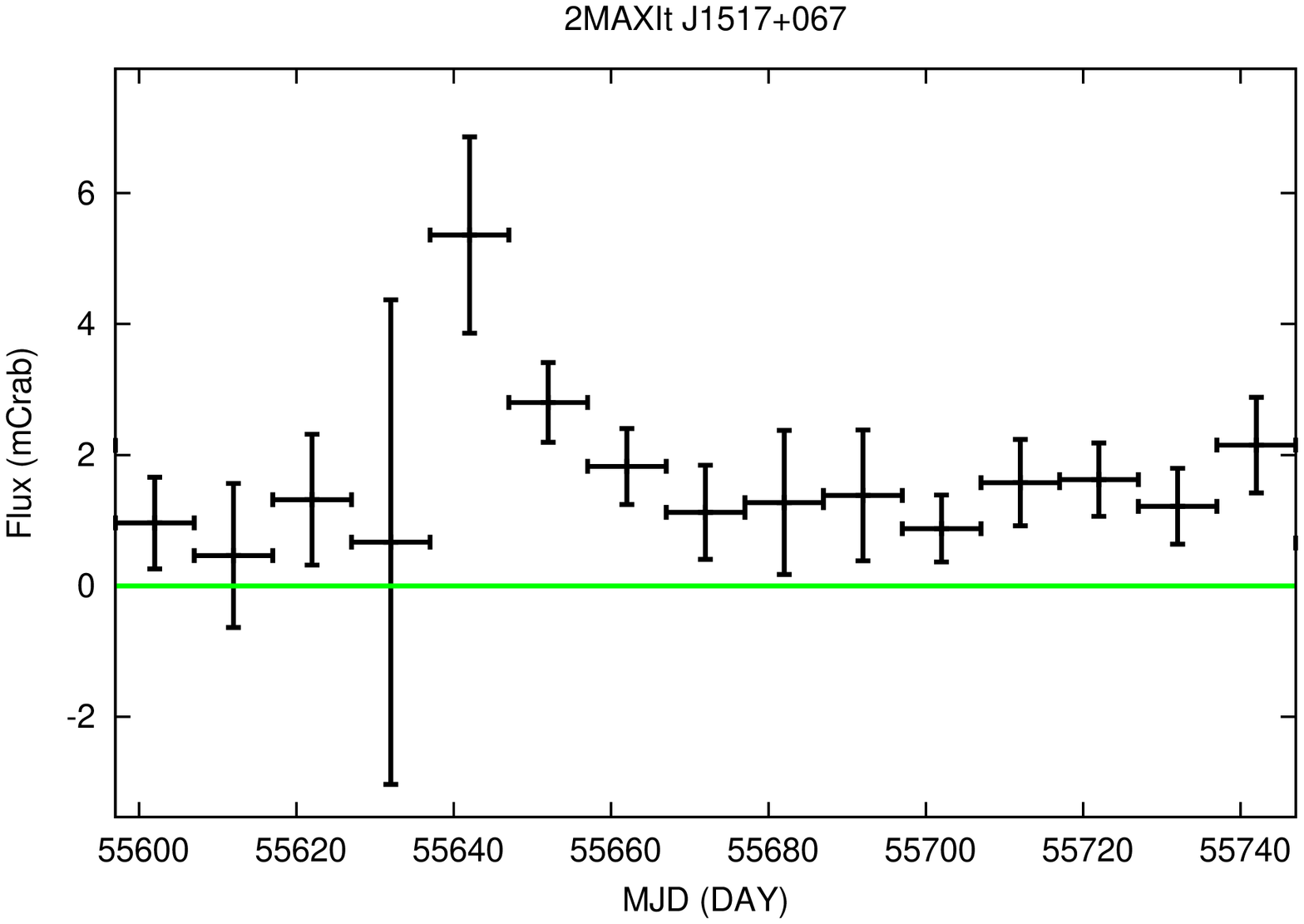}
\includegraphics[scale=0.27,angle=-90]{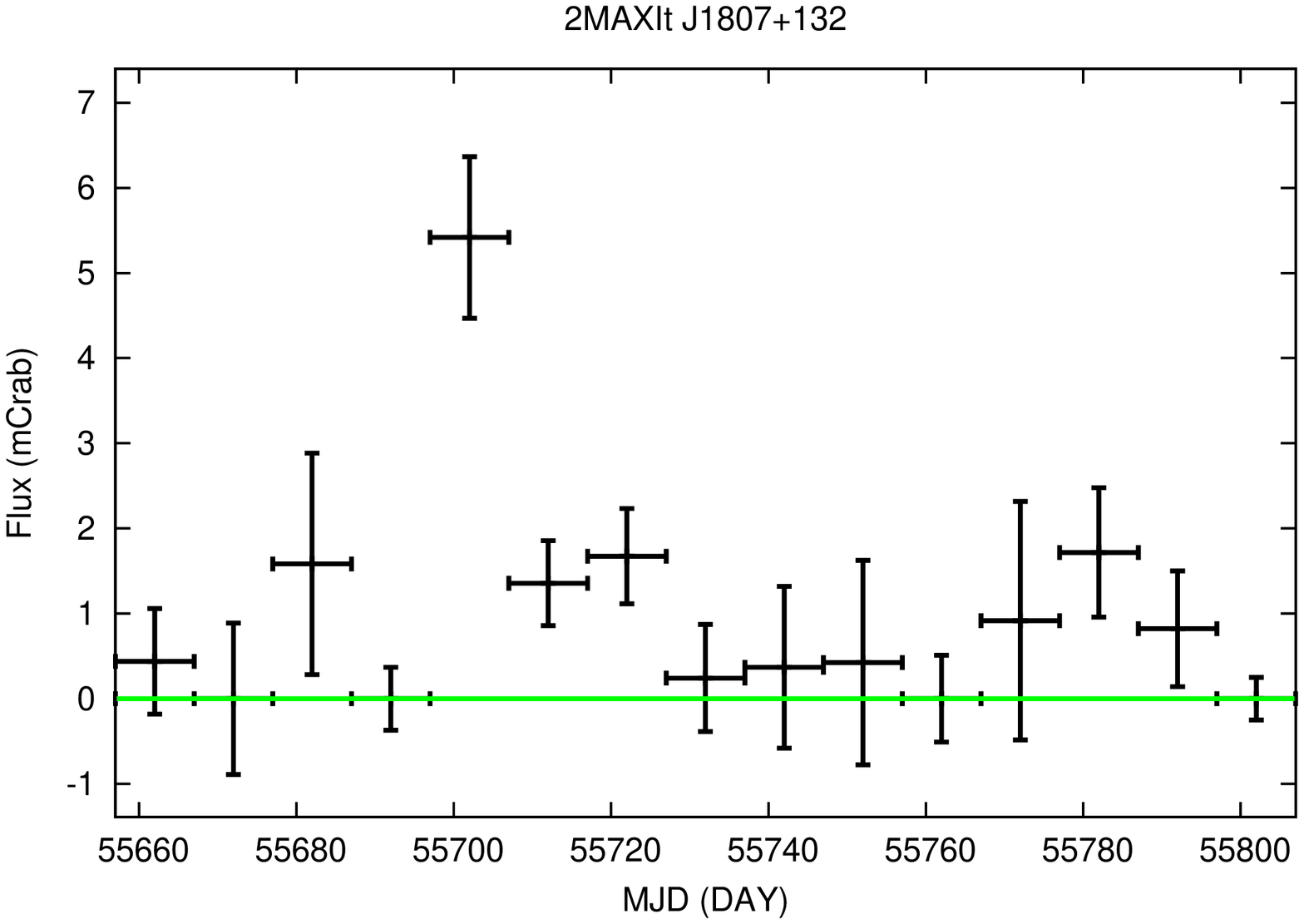}
\includegraphics[scale=0.27,angle=-90]{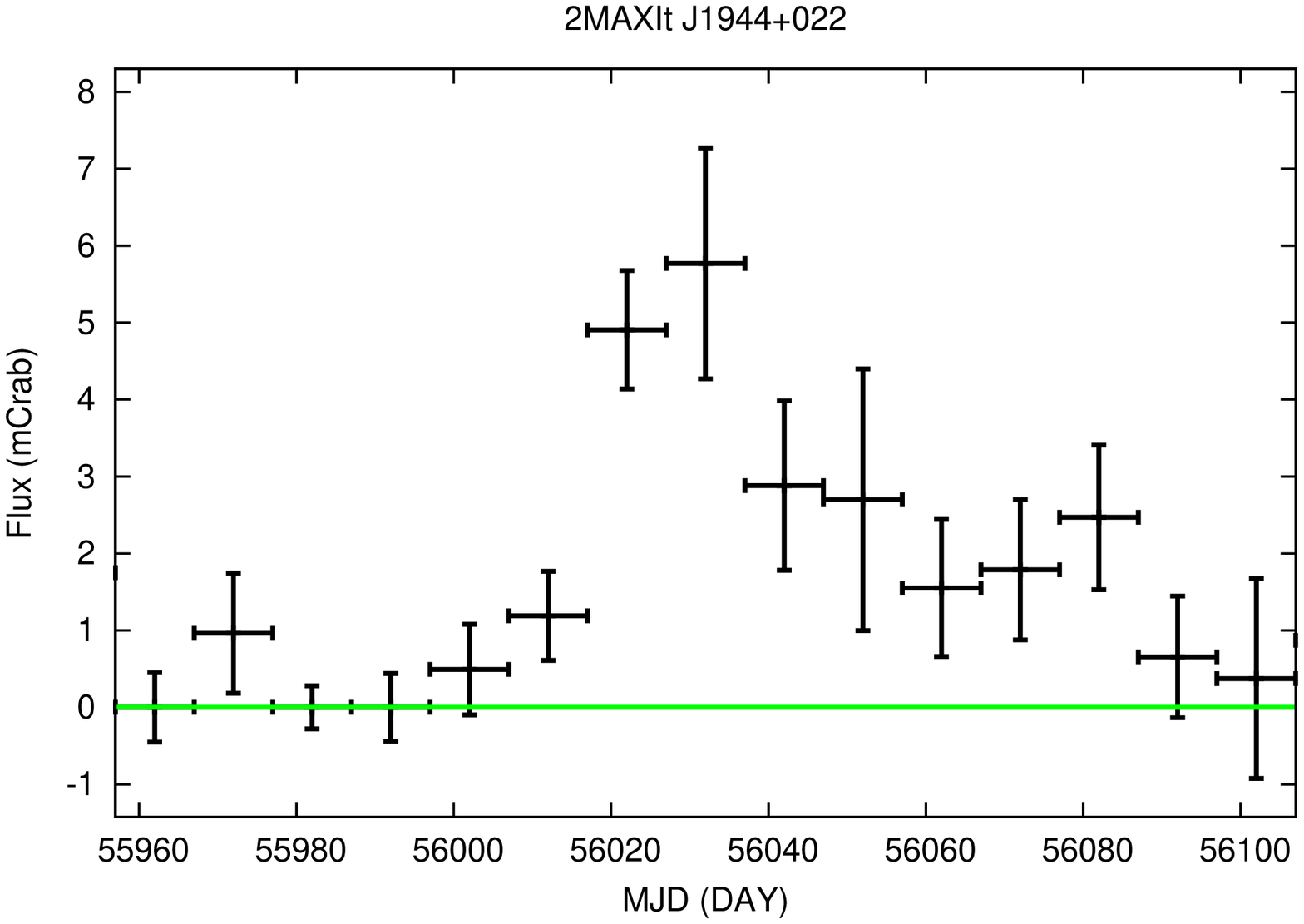}
\includegraphics[scale=0.27,angle=-90]{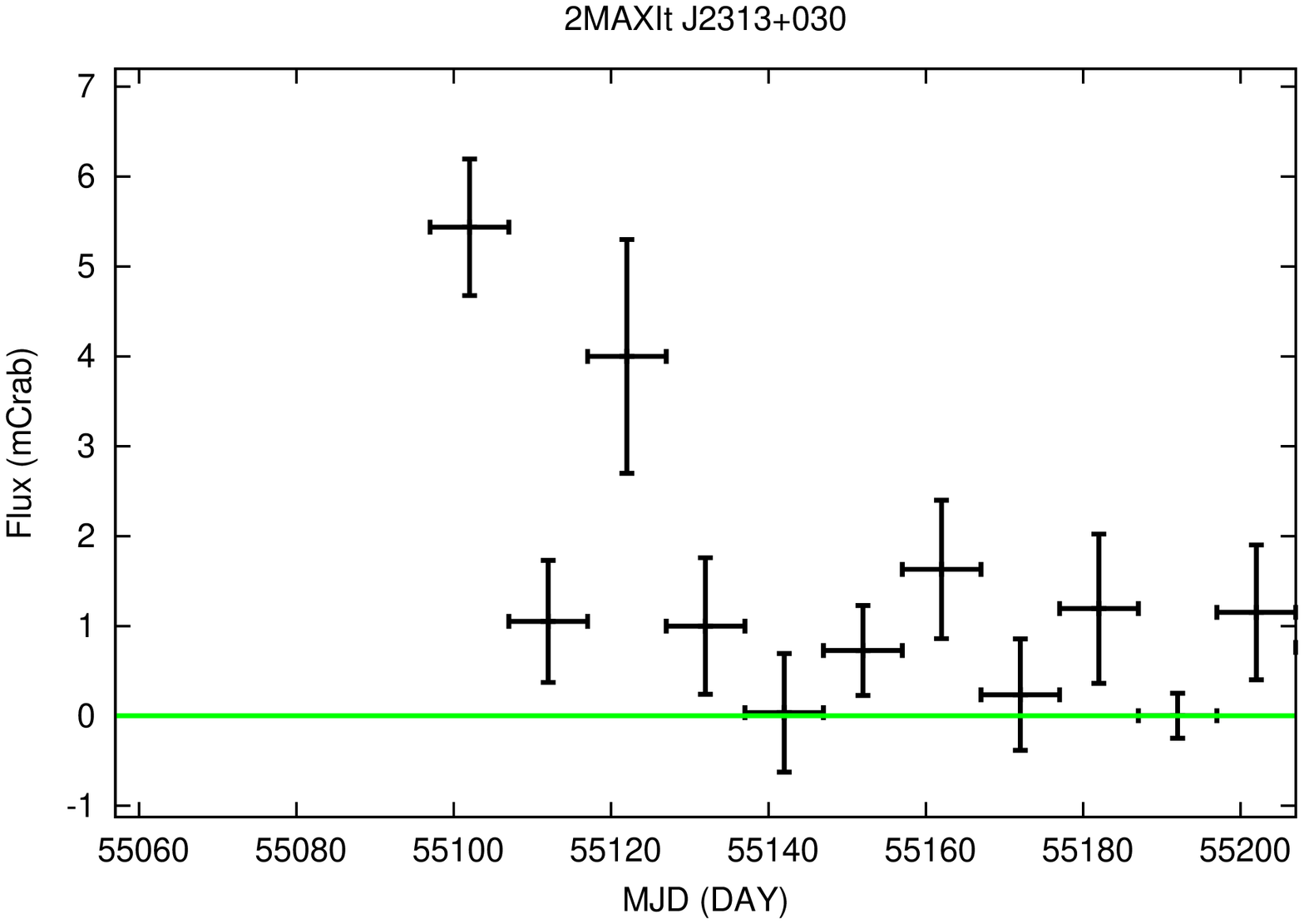}\vspace{0.5cm}
\caption{
The 10-days bin {\it MAXI} light curves in the 3--10 keV band of 
transient sources other than the four TDEs.
}
\label{all_te_lcs}
\end{figure*}

To find possible counterparts of our transient events, we check major
X-ray source catalogs and the literature regarding Gamma-ray Bursts and
TDEs covering our observation epoch: Palermo {\it Swift} BAT X-ray
catalog \citep{Cus10}, {\it Fermi} 2$^{\rm nd}$ LAT Catalog
\citep{Nol12}, {\it ROSAT} Bright Source Catalog \citep{Vog99}, {\it
Swift} BAT 70-month Catalog \citep{Bau13}, {\it Swift} Transient Monitor
Catalog \citep{Kri13}, First {\it XMM-Newton} Sky Slew Survey Catalog
\citep{Sax08}, {\it INTEGRAL} General Reference Catalog (version 36),
and papers by \citet{Cen12}, \citet{Zau13}, \citet{Ser14}, and \citet{Bro15}.  We identify
the counterpart if its position is within the $3\sigma$ positional error
(corresponding to 99\% confidence level) of a {\it MAXI} transient
source.  The error consists of the statistical one and systematic one,
$\sigma_{\rm pos} = (\sigma_{\rm stat}^2 + \sigma_{\rm
sys}^2)^{1/2}$. Here, the systematic error is chosen to be $0^\circ.05$
according to the previous studies (\cite{Hiroi11};
\cite{Hiroi13}). Figure \ref{all_te_lcs} plots the 10-days bin light
curves in the 3--10 keV band of all transient events other than the
three TDEs ({\it Swift} J1112-82,  {\it Swift} J1644+57, and {\it Swift} J2058+05).

\newpage

\clearpage
\begin{landscape}
\begin{table}
\setlength{\topmargin}{10cm}
\caption{Information of Transient Events \label{all_te_info}}
\begin{center}
\begin{tabular}{cccccccccc}
\hline
{\it MAXI} Name & R.A. & Decl. & $\sigma_{\rm pos}$ & $s_{\rm D}$ & $f_{\rm 4-10 keV}$ & Hardness ratio & Flare Time & Counterpart & Type \\
 $[1]$ & $[2]$ & [3]  & [4] & [5] & [6]  & [7] & [8] & [9]  & [10] \\
\hline
2MAXIt J0745$-$504 & 116.451 & -50.496 & 0.216 & 6.07 & 5.16 & $ > $ 0.87 & 55727--55756 & HD 63008 & Star  \\
2MAXIt J1108$-$829 & 167.064 & -82.914 & 0.191 & 6.70 & 2.88 & 0.19 $ \pm $ 0.14 & 55727--55756 & {\it Swift} J1112.2-8238 &  Tidal Disruption Event \\
2MAXIt J1159+238 & 179.814 & 23.876 & 0.185 & 5.77 & 2.66 & $ > $ 0.71 & 55877--55906 &  &   \\
2MAXIt J1507$-$217 & 226.882 & -21.743 & 0.122 & 10.35 & 8.59 & 0.33 $ \pm $ 0.11 & 55127--55156 & GRB 091120 & Gamma-Ray Burst  \\
2MAXIt J1517+067 & 229.350 & 6.793 & 0.198 & 5.79 & 1.79 & 0.23 $ \pm $ 0.18  & 55637--55726 \\ 
2MAXIt J1645+576 & 251.379 & 57.604 & 0.131 & 8.83 & 3.18 & 0.18 $ \pm $ 0.12 & 55637--55726 & {\it Swift} J164449.3+573451 & Tidal Disruption Event  \\
2MAXIt J1807+132 & 271.806 & 13.269 & 0.184 & 5.68 & 2.73 & 0.28 $ \pm $ 0.18 & 55697--55726 &  &   \\
2MAXIt J1944+022 & 296.170 & 2.203 & 0.219 & 6.12 & 2.45 & $ < $ 0.32 & 55997--56086 & {\it Swift} J1943.4+0228 &   CV \\
2MAXIt J2058+053 & 314.578 & 5.377 & 0.202 & 6.39 & 3.39 & 0.19 $ \pm $ 0.15 & 55697--55726 & {\it Swift} J2058.4+0516 & Tidal Disruption Event  \\
2MAXIt J2313+030 & 348.455 & 3.037 & 0.223 & 6.01 & 4.09 & $ < $ 0.23 & 55097--55126 & SZ Psc & RSCVn \\ 
\hline
\multicolumn{1}{@{}l@{}}{\hbox to 0pt{\parbox{150mm}
{\footnotesize
\textbf{Notes.}\\
Col. [1]: {\it MAXI} Name. \\
Col. [2]: Right ascension in units of degree. \\
Col. [3]: Declination in units of degree. \\ 
Col. [4]: 1$\sigma$ statistical position error. \\
Col. [5]: Detection Significance. \\
Col. [6]: Average flux over a time interval in the 4-10 keV band in units of mCrab. \\
Col. [7]: Hardness ratio and its 1$\sigma$ error.\\
Col. [8]: Duration of 30-days bin used when the source is detected.  \\
Col. [9]: Name of the counterpart. \\
Col. [10]: Type of the counterpart. \\
}\hss}}
\end{tabular}
\end{center}
\end{table}

\clearpage
\end{landscape}


\section{{\it MAXI} Results for TDE Candidates Suggested in the {\it Swift}/BAT Data}\label{app:sec:bat_src}

\textcolor{black}{ We check if {\it MAXI} also found any signals of the
TDE candidates reported by \citet{Hry16}, who used the {\it Swift}/BAT
ultra-hard X-ray (20--195 keV) data taken between 2005 and 2013. The
number of TDE candidates detected within the 37-month period of our {\it MAXI}
data is four. We confirm that all of them are not significantly detected
in the 4--10 keV image integrated over a 30 day period covering the
flare time. Thus, it is justified not to include these events in our
analysis. Table~\ref{app:tab:bat_src} summarizes the 5.5$\sigma$ upper
limit of the fluxes in the 4--10 keV band together with the basic
information for each TDE. We find that in all events the {\it MAXI} upper
limit is smaller by a factor of $\sim$4 than that expected from the
averaged BAT flux in the same epoch by assuming a photon index of 2.
The reason is unclear, but these events might be subject to heavy
(nearly Compton thick) obscuration. }

\begin{table*}
\caption{Information of {\it Swift}/BAT TDE Candidates\label{app:tab:bat_src}} 
\begin{center}
\begin{tabular}{ccccc}
\hline
Name & R.A. &  Decl. & $\hat{f}_{\rm 4-10keV}$ & Epoch  \\ 
  $[1]$ & [2] & [3] & [4] & [5] \\  \hline 
PGC 015259  & 67.341  & -4.760 & $<$ 1.7 & 55217--55246 \\ 
UGC 03317   & 83.406  & 73.726 & $<$ 1.1 & 55457--55486 \\
PGC 1185375 & 225.960 &  1.127 & $<$ 3.5 & 55247--55276 \\
PGC 1190358 & 226.370 &  1.293 & $<$ 3.2 & 55247--55276 \\
\hline
\multicolumn{1}{@{}l@{}}{\hbox to 0pt{\parbox{160mm}
{\footnotesize
\textbf{Notes.}\\
Col. [1]: Name of TDE candidates detected with {\it Swift}/BAT \citep{Hry16}.  \\
Col. [2]: Right ascension in units of degree. \\
Col. [3]: Declination in units of degree. \\
Col. [4]: 5.5$\sigma$ upper limit of the averaged 4--10 keV flux in a
 30 day period covering the flare time, in units of mCrab.\\
Col. [5]: Epoch (MJD) when the upper flux limit is estimated. 
}\hss}}
\end{tabular}
\end{center}
\end{table*}


\begin{thebibliography}{1000}


\bibitem[Aharon et al.(2015)]{Aha15} Aharon, D., Mastrobuono Battisti, A., \& Perets, H.~B.\ 2015, arXiv:1507.08287 


\bibitem[Arcavi et al.(2014)]{Arc14} Arcavi, I., Gal-Yam, A., Sullivan, M., et al.\ 2014, \apj, 793, 38 


\bibitem[Bade et al.(1996)]{Bade96} Bade, N., Komossa, S., \& Dahlem, M.\ 1996, \aap, 309, L35 


\bibitem[Baumgartner et al.(2013)]{Bau13} Baumgartner, W.~H., 
Tueller, J., Markwardt, C.~B., et al.\ 2013, \apjs, 207, 19 

\bibitem[Blanton et al.(2001)]{Bla01} Blanton, M.~R., 
Dalcanton, J., Eisenstein, D., et al.\ 2001, \aj, 121, 2358 


\bibitem[Brown et al.(2015)]{Bro15} Brown, G.~C., Levan, 
A.~J., Stanway, E.~R., et al.\ 2015, \mnras, 452, 4297 


\bibitem[Burrows et al.(2011)]{Bur11} Burrows, D.~N., Kennea, 
J.~A., Ghisellini, G., et al.\ 2011, \nat, 476, 421 

\bibitem[Cash(1979)]{Cash79} Cash, W.\ 1979, \apj, 228, 939 

\bibitem[Cenko et al.(2012)]{Cen12} Cenko, S.~B., Krimm, 
H.~A., Horesh, A., et al.\ 2012, \apj, 753, 77 

\bibitem[Chabrier(2003)]{Chab03} Chabrier, G.\ 2003, \pasp, 115, 763 

\bibitem[Cusumano et al.(2010)]{Cus10} Cusumano, G., La Parola, V., Segreto, A., et al.\ 2010, \aap, 524, A64 

\bibitem[Donley et al.(2002)]{Don02} Donley, J.~L., Brandt, 
W.~N., Eracleous, M., \& Boller, T.\ 2002, \aj, 124, 1308 

\bibitem[Eguchi et al.(2009)]{Egu09} Eguchi, S., Hiroi, K., 
Ueda, Y., et al.\ 2009, Astrophysics with All-Sky X-Ray Observations, 44 

\bibitem[Esquej et al.(2007)]{Esq07} Esquej, P., Saxton, R.~D., Freyberg, M.~J., et al.\ 2007, \aap, 462, L49 

\bibitem[Esquej et al.(2008)]{Esq08} Esquej, P., Saxton, R.~D., Komossa, S., et al.\ 2008, \aap, 489, 543 

\bibitem[Evans \& Kochanek(1989)]{Eva89} Evans, C.~R., \& Kochanek, C.~S.\ 1989, \apjl, 346, L13 

\bibitem[Ferrarese(2002)]{Fer02} Ferrarese, L.\ 2002, Current 
High-Energy Emission Around Black Holes, 3 

\bibitem[Gehrels(1986)]{Ge86} Gehrels, N.\ 1986, \apj, 303, 336 



\bibitem[Gezari et al.(2006)]{Gez06} Gezari, S., Martin, 
D.~C., Milliard, B., et al.\ 2006, \apjl, 653, L25 

\bibitem[Gezari et al.(2008)]{Gez08} Gezari, S., Basa, S., 
Martin, D.~C., et al.\ 2008, \apj, 676, 944 


\bibitem[Gezari et al.(2012)]{Gez12} Gezari, S., Chornock, R., Rest, A., et al.\ 2012, \nat, 485, 217 



\bibitem[G{\'o}rski et al.(2005)]{Gor05} G{\'o}rski, K.~M., 
Hivon, E., Banday, A.~J., et al.\ 2005, \apj, 622, 759 

\bibitem[Grupe et al.(1995)]{Gru95} Grupe, D., Beuerman, K., Mannheim, K., et al.\ 1995, \aap, 300, L21 

\bibitem[Grupe et al.(1999)]{Gru99} Grupe, D., Thomas, H.-C., \& Leighly, K.~M.\ 1999, \aap, 350, L31 

\bibitem[Grupe et al.(2015)]{Gru15} Grupe, D., Komossa, S., 
\& Saxton, R.\ 2015, \apjl, 803, L28 

\bibitem[Guillochon \& Ramirez-Ruiz(2013)]{Gui13} Guillochon, J., \& Ramirez-Ruiz, E.\ 2013, \apj, 767, 25 



\bibitem[Hiroi et al.(2011)]{Hiroi11} Hiroi, K., Ueda, Y., Isobe, N., et al.\ 2011, \pasj, 63, 677 

\bibitem[Hiroi et al.(2013)]{Hiroi13} Hiroi, K., Ueda, Y., Hayashida, M., et al.\ 2013, \apjs, 207, 36 

\bibitem[Holoien et al.(2014)]{Hol14} Holoien, T.~W.-S., Prieto, J.~L., Bersier, D., et al.\ 2014, \mnras, 445, 3263 

\bibitem[Hryniewicz \& Walter(2016)]{Hry16} Hryniewicz, K., \& Walter, R.\ 2016, \aap, 586, A9 

\bibitem[Irwin et al.(2015)]{Irw15} Irwin, J.~A., Henriksen, R.~N., Krause, M., et al.\ 2015, \apj, 809, 172 


\bibitem[Isobe et al.(2015)]{Iso15} Isobe, N., Sato, R., Ueda, Y., et al.\ 2015, \apj, 798, 27 

\bibitem[Kippenhahn \& Weigert(1990)]{Kipp90} Kippenhahn, R., \& Weigert, A.\ 1990, Stellar Structure and Evolution, XVI, 468 pp.~192 figs..~ Springer-Verlag Berlin Heidelberg New York.~Also Astronomy and Astrophysics Library,  

\bibitem[Komossa \& Bade(1999)]{Kom99} Komossa, S., \& Bade, N.\ 1999, \aap, 343, 775 

\bibitem[Komossa(2012)]{Kom12} Komossa, S.\ 2012, European 
Physical Journal Web of Conferences, 39, 2001 


\bibitem[Krimm et al.(2013)]{Kri13} Krimm, H.~A., Holland, S.~T., Corbet, R.~H.~D., et al.\ 2013, \apjs, 209, 14 

\bibitem[Kroupa(2001)]{Kro01} Kroupa, P.\ 2001, \mnras, 322, 231 


\bibitem[Li et al.(2002)]{Li02} 
Li, L.-X., Narayan, R., \& Menou, K.\ 2002, \apj, 576, 753 


\bibitem[Magorrian \& Tremaine(1999)]{Mag99} Magorrian, J., \& Tremaine, S.\ 1999, \mnras, 309, 447 

\bibitem[Maksym et al.(2010)]{Mak10} Maksym, W.~P., Ulmer, M.~P., \& Eracleous, M.\ 2010, \apj, 722, 1035 

\bibitem[Marconi \& Hunt(2003)]{Mar03} Marconi, A., \& Hunt, L.~K.\ 2003, \apjl, 589, L21 


\bibitem[Marconi et al.(2004)]{Mar04} Marconi, A., Risaliti, G., Gilli, R., et al.\ 2004, \mnras, 351, 169 

\bibitem[Matsuoka et al.(2009)]{Mat09} Matsuoka, M., 
Kawasaki, K., Ueno, S., et al.\ 2009, \pasj, 61, 999 

\bibitem[Mihara et al.(2011)]{Mih11} Mihara, T., Nakajima, 
M., Sugizaki, M., et al.\ 2011, \pasj, 63, 623 

\bibitem[Milosavljevi{\'c} et al.(2006)]{Milo06} 
Milosavljevi{\'c}, M., Merritt, D., \& Ho, L.~C.\ 2006, \apj, 652, 120 


\bibitem[Miyaji et al.(2001)]{Miya01} Miyaji, T., Hasinger, G., \& Schmidt, M.\ 2001, \aap, 369, 49 


\bibitem[Negoro et al.(2012)]{Neg12} Negoro, H., Ozawa, H., 
Suwa, F., et al.\ 2012, Astronomical Data Analysis Software and Systems 
XXI, 461, 797 


\bibitem[Niko{\l}ajuk \& Walter(2013)]{Niko13} Niko{\l}ajuk, M., \& Walter, R.\ 2013, \aap, 552, A75 


\bibitem[Nolan et al.(2012)]{Nol12} Nolan, P.~L., Abdo, A.~A., Ackermann, M., et al.\ 2012, \apjs, 199, 31 

\bibitem[Pasham et al.(2015)]{Pas15} Pasham, D.~R., Cenko, S.~B., Levan, A.~J., et al.\ 2015, \apj, 805, 68 

\bibitem[P{\'e}rez-Gonz{\'a}lez et al.(2005)]{Per05} 
P{\'e}rez-Gonz{\'a}lez, P.~G., Rieke, G.~H., Egami, E., et al.\ 2005, \apj, 630, 82 

\bibitem[Phinney(1989)]{Phi89} Phinney, E.~S.\ 1989, The Center of the Galaxy, 136, 543 

\bibitem[Press et al.(1992)]{Pre92} Press, W.~H., Teukolsky, 
S.~A., Vetterling, W.~T., 
\& Flannery, B.~P.\ 1992, Cambridge: University Press, |c1992, 2nd ed.,  

\bibitem[Rees(1988)]{Rees88} Rees, M.~J.\ 1988, \nat, 333, 523 


\bibitem[Salpeter(1955)]{Sal55} Salpeter, E.~E.\ 1955, \apj, 
121, 161 

\bibitem[Saxton et al.(2008)]{Sax08} 
  Saxton, R.~D., Read, A.~M., Esquej, P., et al.\ 2008, \aap, 480, 611 

\bibitem[Saxton et al.(2012)]{Sax12} Saxton, R.~D., Read, A.~M., Esquej, P., et al.\ 2012, \aap, 541, A106 

\bibitem[Serino et al.(2014)]{Ser14} Serino, M., Sakamoto, 
T., Kawai, N., et al.\ 2014, arXiv:1406.3912 

\bibitem[Shankar et al.(2004)]{Sha04} Shankar, F., Salucci, 
P., Granato, G.~L., De Zotti, G., \& Danese, L.\ 2004, \mnras, 354, 1020 


\bibitem[Soltan(1982)]{Sol82} Soltan, A.\ 1982, \mnras, 200, 115 

\bibitem[Stone \& Loeb(2012)]{Sto12} Stone, N., \& Loeb, A.\ 2012, Physical Review Letters, 108, 061302 

\bibitem[Stone \& Metzger(2014)]{Sto14} Stone, N.~C., \& Metzger, B.~D.\ 2014, arXiv:1410.7772 

\bibitem[Sugizaki et al.(2011)]{Sug11} Sugizaki, M., Mihara, 
T., Serino, M., et al.\ 2011, \pasj, 63, 635 


\bibitem[Tomida et al.(2011)]{Tom11} Tomida, H., Tsunemi, H., 
Kimura, M., et al.\ 2011, \pasj, 63, 397 


\bibitem[Ueda et al.(2014)]{Ueda14} Ueda, Y., Akiyama, M., 
Hasinger, G., Miyaji, T., \& Watson, M.~G.\ 2014, \apj, 786, 104 


\bibitem[Ulmer(1999)]{Ulm99} Ulmer, A.\ 1999, \apj, 514, 180 

\bibitem[van Velzen et al.(2011)]{Van11} van Velzen, S., 
Farrar, G.~R., Gezari, S., et al.\ 2011, \apj, 741, 73 

\bibitem[van Velzen \& Farrar(2014)]{Van14} van Velzen, S., \& Farrar, G.~R.\ 2014, \apj, 792, 53 

\bibitem[Vasudevan \& Fabian(2007)]{Vas07} Vasudevan, R.~V., \& Fabian, A.~C.\ 2007, \mnras, 381, 1235 


\bibitem[Voges et al.(1999)]{Vog99} Voges, W., Aschenbach, B., Boller, T., et al.\ 1999, \aap, 349, 389 

\bibitem[Wang \& Merritt(2004)]{Wang04} Wang, J., \& Merritt, D.\ 2004, \apj, 600, 149 


\bibitem[Zauderer et al.(2011)]{Zau11} Zauderer, B.~A., 
Berger, E., Soderberg, A.~M., et al.\ 2011, \nat, 476, 425 

\bibitem[Zauderer et al.(2013)]{Zau13} Zauderer, B.~A., Berger, E., Margutti, R., et al.\ 2013, \apj, 767, 152 

\end{thebibliography}
\end{document}